\documentclass[]{aastex631}

\begin{document}

\title{Spectroscopic Links Among Giant Planet Irregular Satellites and Trojans}

\author[0000-0003-1383-1578]{Benjamin N. L. Sharkey}
\affiliation{Department of Astronomy, University of Maryland \\
4296 Stadium Dr. \\
PSC (Bldg 415) Rm 1113 \\
College Park, MD 20742-2421, USA \\
}
\affiliation{Lunar and Planetary Laboratory, University of Arizona \\
1629 E University Blvd \\
Tucson, AZ 85721-0092, USA}

\author[0000-0002-7743-3491]{Vishnu Reddy}
\affiliation{Lunar and Planetary Laboratory, University of Arizona \\
1629 E University Blvd \\
Tucson, AZ 85721-0092, USA}

\author[0000-0002-7389-1655]{Olga Kuhn}
\affil{Large Binocular Telescope Observatory\\
University of Arizona\\
933 N. Cherry Ave, Room 552\\
Tucson, AZ 85721, U.S.A.}

\author[0000-0002-0764-4672]{Juan A. Sanchez}
\affiliation{Planetary Science Institute \\
1700 E Fort Lowell Suite 106 \\
Tucson, AZ, 85719, USA}

\author[0000-0002-0764-4672]{William F. Bottke}
\affiliation{Department of Space Studies, Southwest Research Institute \\
1050 Walnut Street, Suite 300 \\ 
Boulder, CO 80302, USA}

\begin{abstract}
We collect near-infrared spectra ($\sim0.75-2.55\ \mu m$) of four Jovian irregular satellites and visible spectra ($\sim0.32-1.00\ \mu m$) of two Jovian irregular satellites, two Uranian irregular satellites, and four Neptune Trojans. We find close similarities between observed Jovian irregular satellites and previously characterized Jovian Trojans. However, irregular satellites' unique collisional histories complicate comparisons to other groups. Laboratory study of CM and CI chondrites show that grain size and regolith packing conditions strongly affect spectra of dark, carbonaceous materials. We hypothesize that different activity histories of these objects, which may have originally contained volatile ices that subsequently sublimated, could cause differences in regolith grain-size or packing properties and therefore drive spectral variation. The Uranian satellites Sycorax and Caliban appear similar to TNOs. However, we detect a feature near 0.7 $\mu m$ on Sycorax, suggesting the presence of hydrated materials. While the sample of Neptune Trojans have more neutral spectra than the Uranian satellites we observe, they remain consistent with the broad color distribution of the Kuiper belt. We detect a possible feature near 0.65-0.70 $\mu m$ on Neptune Trojan 2006 RJ103, suggesting that hydrated material may also be present in this population. Characterizing hydrated materials in the outer solar system may provide critical context regarding the origins of hydrated CI and CM chondrite meteorites. We discuss how the hydration state(s) of the irregular satellites constrains the thermal histories of the interiors of their parent bodies, which may have formed among the primordial Kuiper belt.
\end{abstract}

\section{Introduction} \label{sec:c5_intro}

The irregular satellites of the giant planets are small, loosely bound objects in a variety of highly eccentric and inclined orbits about Jupiter, Saturn, Uranus, and Neptune. With the exception of Triton, these satellites are all less than 500 km in diameter, with the vast majority of known objects smaller than 100 km. These satellite swarms are readily understood as objects captured from previously heliocentric orbits. However, their provenance and the exact timing of their capture are unknown. Few irregular satellites have been compositionally characterized, leaving questions regarding their origins unresolved. But forging connections between irregular satellites and their heliocentric parent population(s) is critical to assess how primitive material was transported within the giant planet region, as well as the dynamical histories of the giant planets themselves.

Various scenarios have been put forward to explain the origins of irregular satellites, such as pull-down capture \citep{Heppenheimer1977}, gas-drag \citep{Pollack1979,Astakhov2003,Cuk2004}, and three-body interactions \citep{Colombo1971,Nesvorny2007,Nesvorny2014}. These scenarios can be divided into two hypotheses: local or nonlocal. Generally, pull-down capture and gas-drag capture would emplace irregular satellite systems early, during the formation of the host planet. Early capture means the irregular satellites are samples of material close to the forming planet itself (capture from ``local'' sources). Meanwhile, three body capture can occur early (also from ``local'' sources) or much later, during hypothesized encounters between giant planets during an instability \citep[e.g.,][]{Nesvorny2007}. Any original captured populations (including those emplaced via pull-down or gas-drag) would be depleted by planetary encounters before being replenished from whatever background planetesimal populations exist at that time (capture from ``nonlocal'' sources). Detailed reviews of origin scenarios are given by \citet{Nesvorny2018}, \citet{Jewitt2007}, and \citet{Nicholson2008}.

Dynamical models of giant planet migration suggest that irregular satellites began their lives in the primordial Kuiper belt (PKB), a region located just beyond the original formation location of the giant planets between $\sim5$ and $\sim17$ au \citep[e.g.,][]{Tsiganis2005,Nesvorny2012,Nesvorny2019}. The PKB is thought to have stretched between $\sim24$ and $\sim50$ au, with an initial mass that was the order of several tens of Earth masses \citep{Nesvorny2019}.

Modeling work indicates that the PKB may have lasted for several tens of Myr after the dispersion of the solar nebula, but no longer than 100 Myr \citep{Nesvorny2018b}. The end of the PKB was brought about when Neptune migrated across it, thereby ejecting 99.9\% of the PKB population onto giant planet-crossing orbits.  We will refer to this component as the destabilized population.  Most of it was handed down to Jupiter by planetary close encounters, but a smaller component was scattered outward by Neptune, where it now comprises the scattered disk. Neptune's migration also triggered a dynamic instability among the giant planets \citep{Tsiganis2005}. This led the giant planets to have encounters with one another while being surrounded by members of the destabilized population. The combination led to three-body reactions that captured a small fraction of this destabilized population into stable zones surrounding the giant planets \citep{Nesvorny2007}. \citet{Nesvorny2014} demonstrate that the number and orbital properties of the irregular satellite systems can be well-matched via these sequences of planetary encounters.

Understanding if the irregular satellites of Jupiter, Saturn, Uranus, and Neptune share similar physical properties is an open question that directly tests whether they share origins from the PKB, as predicted by these large-scale migration hypotheses. Additional comparisons can be made to other populations expected to share this lineage, like Trojans, excited TNOs, and Centaurs. Complications can arise in cross-population comparisons, however, due to the possibility for divergent evolutionary histories after their emplacement. 

For the irregular satellites, collisions may be a uniquely important element of their post-capture evolution. The reasons are twofold – the volume of space taken up by the irregulars around each giant planet is relatively small, while their impact velocities with one another  are relatively high (e.g., the order of many km/s; see Table 2 in \citet{Bottke2010}). Collisional simulations indicate most irregular objects were decimated in short order, such that only a small remnant population has lasted to today \citep{Bottke2010}. In this scenario, most existing irregulars are likely to be fragments produced by or survivors shaped by collisions. A few of the largest irregulars may still have much of their original mass, but they were likely shattered by impact events. \citet{Bottke2010} also demonstrates that the irregular satellites' size distribution can be well-produced from intense collisional grinding of an original population that was similar to that of the present-day Jovian Trojans. This suggests that collisions are likely one of the dominant processing mechanisms for irregular satellites. This collisional grinding would produce debris, and the presence of dark, carbonaceous material on the exteriors of the Galilean satellites \citep{Bottke2013} and the large satellites of Uranus \citep{cartwright2018} provides observable evidence suggestive of this process. 

A further implication of \citet{Bottke2010} is that the Jovian Trojans maintained a size distribution that was minimally altered after their capture. Considering that the capture of Jovian Trojans can also be explained via planetary encounters \citep{Nesvorny2013}, this line of evidence points towards Jovian Trojans and irregular satellites sharing similar origins. Therefore, learning how consistent these populations' present-day compositions are provides a strict test of these migration and evolutionary models. While less well-studied than the Jovian system, Neptune's Trojans and irregular satellite system can also be explored under this lens.

Observational evidence has been brought forward to strengthen the specific connection between the Jovian Trojans and outer solar system material, e.g., the hypothesis of \citet{Wong2016} that links the color bimodality of the Jovian Trojans with similar, but distinct, color bimodalities found among Centaurs and small Kuiper Belt objects. Under this view, the present-day Jupiter Trojans are understood as bearing the modified record of two surface types whose origins can still be found in the trans-Neptunian region. Other results draw an affinity between some Trojans and hydrated, irradiated material similar to CM2 carbonaceous chondrites \citep{Yang2011}. Assessing the origins of the Jovian Trojans drives NASA's recently launched Lucy mission \citep{Levison2021}. Understanding how irregular satellite surfaces vary with heliocentric distance can provide crucial insight into questions regarding the evolution of small body surfaces in the outer solar system.

\subsection{Current Observational Knowledge}

There are no clear, population-wide compositional links between Trojans and irregular satellites at present. This is largely due to the fundamental ambiguities of interpreting the spectra of dark, carbonaceous C/D/P-type asteroids that lack mineralogically diagnostic absorption features. \citet{Bhatt2017} conducted a spectroscopic survey of the Jovian satellites Himalia, Elara, and Carme, finding that the reflectance properties of these objects can be well matched by virtual linear mixtures of hydrated silicates and carbonaceous materials, analogous to the compositions of hydrated carbonaceous chondrites. \citet{Brown2014} discovered clear evidence for hydration amongst the Jovian satellites, finding that Himalia has a 3-$\mu m$ absorption feature (indicative of hydrated materials) similar to the main-belt asteroid (52) Europa \citep{Takir2012}.

The best explored small irregular satellite of any system, Phoebe, was observed by the Cassini spacecraft and found to contain a variety of hydrated surface materials and organics, linking this satellite to an origin in the outer solar system \citep{Clark2005}. Further analysis showed that Phoebe's water ice content is consistent with observations of dynamically excited KBOs \citep{Fraser2018}. The irregular satellites of Jupiter, Saturn, Uranus, and Neptune have been observed via visible and near-infrared color photometry \citep{Graykowski2018,Grav2003,Grav2007,Rettig2001,Sykes2000}. \citet{Vilas2006} conducted a spectrophotometric search for the 0.7 micron absorption feature (which is linked to hydrated materials) in several irregular satellite systems, finding evidence for this feature on several Jovian satellites, the Saturnian satellite Phoebe, and the Uranian satellite Caliban. \citet{Vilas2022} conducted a narrow-band spectroscopic search near 0.7 microns and suggest that various dynamical clusters may represent materials excavated from different depths by catastrophic collisions of their original parent bodies.

\citet{Graykowski2018} collect color observations of each of the irregular satellite populations, and by compiling their observations with previous literature values, find that the color properties of all four populations overlap. Notably, they find that the irregular satellites lack ultra-red material seen in the Kuiper Belt. If the irregular satellites were captured from the primordial Kuiper Belt, this observational evidence suggests a modification process to remove ultra-red surfaces is necessary to explain their present-day properties. Due to the wide range of surface temperatures between Jupiter and Neptune, they conclude that thermal alteration effects are unlikely to uniformly explain this discrepancy.

No spectroscopic measurements have been reported for the Uranian irregular satellites. At Neptune, excluding Triton, only Nereid has been spectroscopically observed, and it has been found to have abundant water ice with neutral visible to near-infrared colors \citep{mebrown1998,rhbrown1999,mebrown2000,Sharkey2021a}. Recent NIR spectra collected for several Neptune Trojans find compositional distinctions between red and ultra-red members of this population \citep{Markwardt2023}.

 Understanding the compositional variability of these systems is necessary to understand their origins. Translating irregular satellite and Trojan color properties to compositional information is therefore critical. The work reported here targets objects spanning a range of sizes within each population and explores how their spectral variation can be interpreted. This spectral analysis serves as a step towards establishing compositional linkages or distinctions amongst the giant planet region, which ultimately traces the complexity of small body migration and evolution in the outer solar system.

\section{Observations and Data Reduction} \label{sec:c5_obsdata}

A summary of observations is given in Tables \ref{tab:jov_sats_tab} and \ref{tab:icegiant_obj_table}. Table \ref{tab:jov_sats_tab} describes the observational circumstances of Jovian satellites with the NASA Infrared Telescope Facility (IRTF) in 2018 and 2021 at near-infrared wavelengths ($\sim0.7-2.52 \mu m$), and the Large Binocular Telescope (LBT) in 2018 and 2019 at visible wavelengths ($0.32-1.00 \mu m$). Jovian targets range in brightness from $V \sim 18-21$. Similarly, Table \ref{tab:icegiant_obj_table} describes observations via LBT of four Neptune Trojans and the two largest Uranian irregular satellites, Sycorax and Caliban. These outer solar system objects range in brightness from $V \sim 20-23$. All targets were observed at phase angles of $\sim10^{\circ}$ or less. Table \ref{tab:icegiant_obj_table} indicates both local standard stars used to correct for telluric distortions and solar analog stars used to convert arbitrary flux into reflectance values. This method derives relative reflectance as the ratio of instrumental flux from the target to the solar analog star, and does not fully derive and correct the instrumental response to calibrate spectra to absolute fluxes using a spectrophotometric standard.

Infrared observations were collected via the SpeX instrument on the IRTF using a 0.8'' slit in prism mode. Corrections to instrumental flux were made by measuring a nearby G-type star to correct for telluric absorption features approximately once per hour. Independent observations of the solar analog star SAO 120107 were made each observing night to convert flux to reflectance. Data were processed using Spextool \citep{cushing2004}, and our procedures follow the methodologies of \citet{sanchez2013} and \citet{sanchez2015}. The standard photometric uncertainties output via Spextool's processing are displayed as the error bars in all figures displaying SpeX infrared spectra. The infrared spectra were subsequently mean-binned by a factor of five weighted by the uncertainties derived from Spextool. This was performed to aid the legibility of our plots which compare spectra of multiple objects, and does not affect our analysis or conclusions compared to unbinned data.

Observations at LBT were performed with the twinned Multi-Object Double Spectrograph (MODS) visible spectrographs in long-slit mode \citep{Pogge2010}. The MODS2 red channel was offline during the campaigns in 2018 due to a problem with the head electronics, and therefore MODS2 was used in blue-channel-only mode while both red and blue channels of MODS1 were used. Observations of Jovian satellite Leda were conducted with a 2.4'' wide slit, while all other observations were performed with a 1.2'' wide slit. On all observations, the slit position angle was aligned within $10^{\circ}$ of the parallactic angle. After performing initial image processing using the modsCCDRed suite of scripts to subtract the bias, flat-field and correct for bad pixels, and create 2D wavelength maps from the comparison lamp spectra,  MODS spectra were reduced using custom routines written in python, with a similar approach as in \citet{Sharkey2021b}. This includes an implementation of the optimal extraction algorithm described by \citet{Horne1986}, with sky subtraction adapted from the flame pipeline \citep{Belli2018} which makes use of the method of \citet{Kelson2003}. We provide additional details of this processing routine here for clarity.

Individual integrations were first median stacked before fitting and subtracting the sky based on regions adjacent to the target trace. The method of \citet{Kelson2003} preserves the original on-chip image orientation (i.e., the wavelength axis is tilted with respect to the XY image coordinates). Two-dimensional wavelength maps of the slit image were used to produce a one-dimensional b-spline model of sky background flux as a function of wavelength. The sky model was fit based on local regions of the sky near the target trace, and manually inspected to ensure sky residuals were consistent with zero (within noise).

The extraction profiles, which apply as weights to the algorithm of \citet{Horne1986}, were created by constructing a spatial profile of the bright solar analog star as a function of wavelength. The profiles were created by applying a moving-average filter across the sky-subtracted spectral image of the star's trace, then normalizing each column in the profile by the total stellar flux per column, such that the sum of each column in the profile is equal to one. These extraction profiles were applied to the target spectra by first fitting the centers of the faint target trace with a low-order polynomial, then translating the extraction profile to this location in pixel coordinates. This process was performed interactively to minimize errors introduced to the spectral continuum by imprecise tracing of our faint sources. This method is computationally inexpensive given the size of our data set and is useful for rapid reduction of longslit observations of individual sources.

Due to the relatively high spectral resolution for our purposes $(R \sim 2000)$, and the low per-pixel SNR of our observing mode, we median-binned the extracted one-dimensional spectrum after performing the sky subtraction and telluric corrections. Binning factors are arbitrary, but were chosen to be 600 pixels for the blue channel and 300 pixels for the red channel to produce spectra with approximately consistent SNR between the two modes. The partial (red-only) observations of 2010 TS191 and 2010 TT191 are the exception to this choice, with spectra binned to a coarser width of 425 pixels. Such binning does not affect our efforts to detect and/or characterize the broad crystal field absorption features we seek, which are typical of solid asteroidal surfaces, but does preclude searches for narrower features. Flux errors were taken as the standard error of the median within each bin and are represented by the error bars in all figures displaying MODS visible spectra. The choice of using unweighted median statistics in our binning procedure seeks to maximally suppress imperfections of the sky background subtraction. Such imperfections were noted particularly from the presence of bright, narrow telluric emission from OH lines at wavelengths greater than $\sim 0.7 \mu m$. An extinction correction based on the mean observed airmass for each target was applied to standard star and target spectra to normalize fluxes to unit airmass before deriving relative reflectance as the ratio of target flux to solar analog star flux. We note that, due to the faintness of our targets, long integrations of $\sim 30$ minutes meant that the airmass of the target changed appreciably during each exposure (as given by the airmass ranges in Table \ref{tab:icegiant_obj_table}, which give the range of airmasses encompassed by the start of the first and the start of the last integration). 

It is worth investigating variations resulting from assuming a single airmass value for the extinction correction (required since spectra are co-added before extraction to recover the trace). This was performed by varying the extinction correction across the full airmass range of each target. We find variations of $\sim 10\%$ about the value derived from using the median airmass value. While this is an overestimate of the error introduced in this step (since most observations are closer to the median airmass than not), it provides a reasonable assessment of this source of systematic error.

Observations in the red and blue channels of the MODS instrument are processed and treated separately. The scaling between the blue and red channel data is preserved, and except where noted, is not arbitrary.

To assess the consistency of our data, we must account for variations imposed by telluric emission features, particularly in the wavelength range of 0.8-1.0 microns. We use our observations of Themisto, presented in Figure \ref{fig:c5_fig1}, which were conducted using both pairs of the MODS blue and red instruments, to provide a systematic assessment of our methodology. At the red end, scatter of $\sim 5\%$ is evident between the two instruments, which we interpret as the uncertainty limit from our sky subtraction (as this region is near many sky emission lines). Agreement from $0.55-0.80 \mu m$ is excellent (within uncertainty) between the two instruments.

Observations of Jovian satellite Leda were interrupted due to high winds before the solar analog could be observed. Instead, spectrophotometric observations of the flux standard star Feige 34 (spectral type sdOp D) were used to convert flux to relative reflectance. This was performed by dividing the target flux by that of Feige 34, then computing the solar correction as the ratio of the spectral energy distributions of Feige 34 to the reference solar spectrum of the CALSPEC database \citep{Bohlin2014}.

\section{Results} \label{sec:c5_results}

\subsection{Jovian System}

\begin{figure}[h!]
\includegraphics[width=\textwidth]{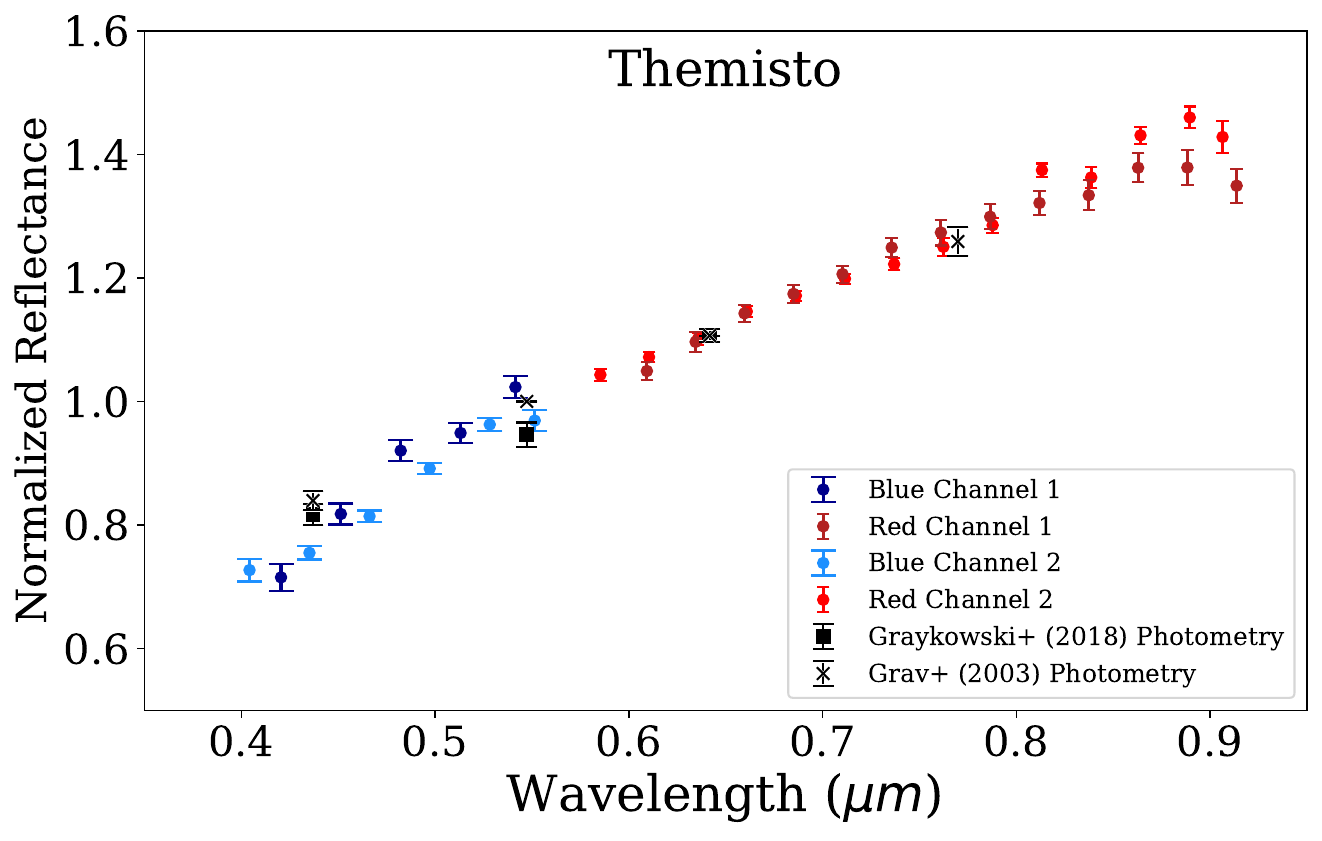}
\includegraphics[width=\textwidth]{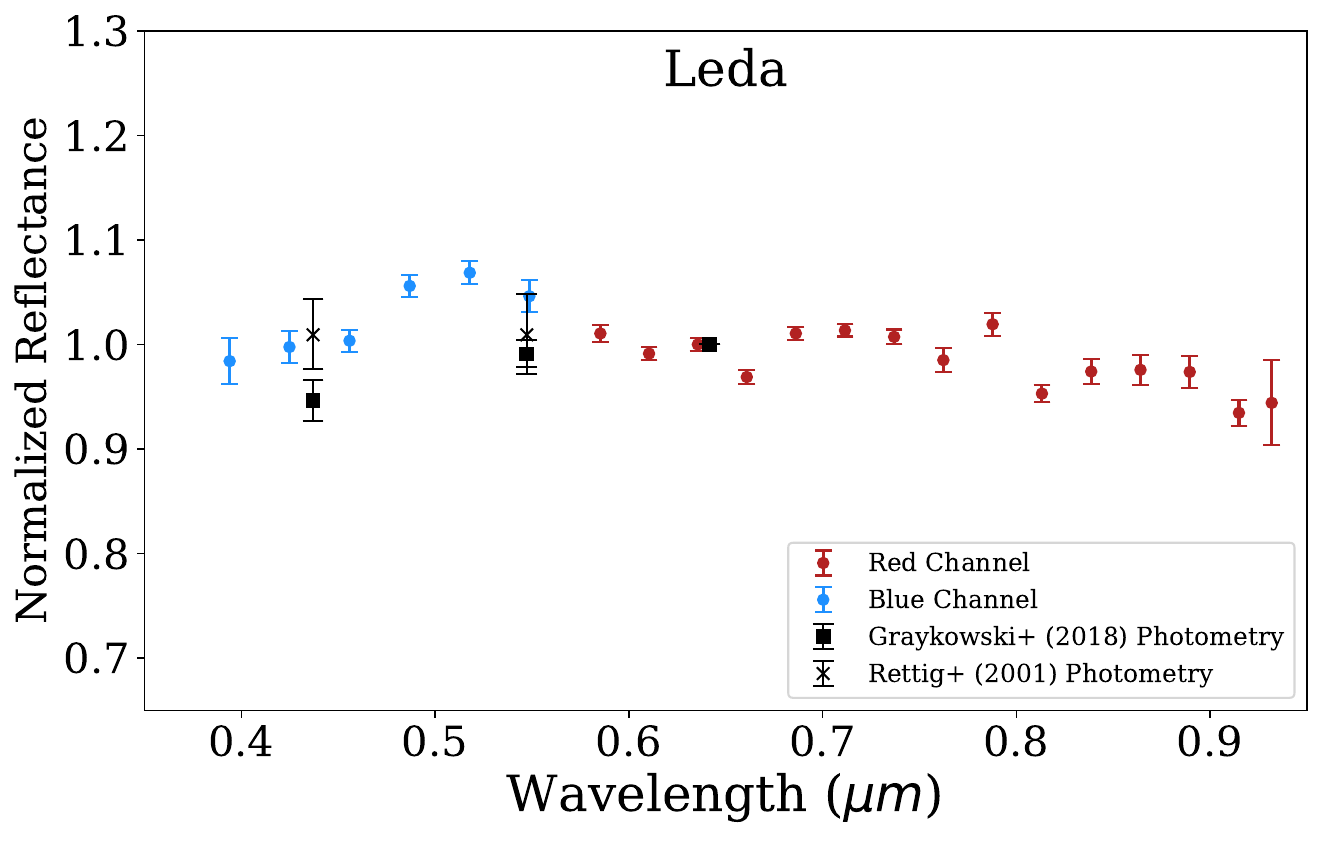}
\caption{Jovian moon spectra from this work plotted with photometric BVR colors from the literature, normalized to unity at the central wavelength of the R filter. We note strong overlap between the spectra derived between the red and blue channels on both of the MODS instruments for Themisto. Leda was observed with MODS1 Red and MODS2 blue channels.}
\label{fig:c5_fig1}
\end{figure}

The visible spectra of the faint Jovian irregular satellites Themisto and Leda are presented in Figure \ref{fig:c5_fig1}. Figure \ref{fig:c5_fig1} also presents comparisons to literature color values as taken from \citet{Graykowski2018}, \citet{Grav2003}, and \citet{Rettig2001}). We find Themisto to display a steep, D-type-like spectrum in good agreement with both \citet{Graykowski2018} and \citet{Grav2003}. The spectra collected of Themisto display good agreement (in both red and blue channels) between MODS1 and MODS2. Spectral slope calculations are also given in Table \ref{tab:jov_sats_tab} as a quantitative reference for each object's spectral behavior, though we caution that such measurements assume perfectly linear slopes across the given baselines.

We find that Leda displays a visible spectral slope similar to those reported by photometric observations, with a rise in slope from $\sim 0.4-0.5 \mu m$ before rolling over to a blue slope over the range of $\sim 0.55-0.90 \mu m$, similar to visible photometric measurements \citep{Rettig2001,Graykowski2018}. This spectrum is also similar to the colors of Himalia, the largest irregular satellite in Leda's dynamical family, as reported by \citet{Jarvis2000}. Similarities between Leda and Himalia were previously noted from infrared photometry \citep{Grav2004a}. We report Leda's spectrum to display curvature from $0.6-0.9 \mu m$, although the achieved SNR limits detailed searches for faint absorption features. Such features may be expected, however, as the presence of a $0.7 \mu m$ absorption feature for Himalia was reported by \citet{Jarvis2000} and is consistent with an understanding of hydrated silicate material in the Himalia family, as shown by the detection of a $3.0 \mu m$ feature on Himalia by \citet{Brown2014}.

\begin{figure}[h!]
\includegraphics[width=\textwidth]{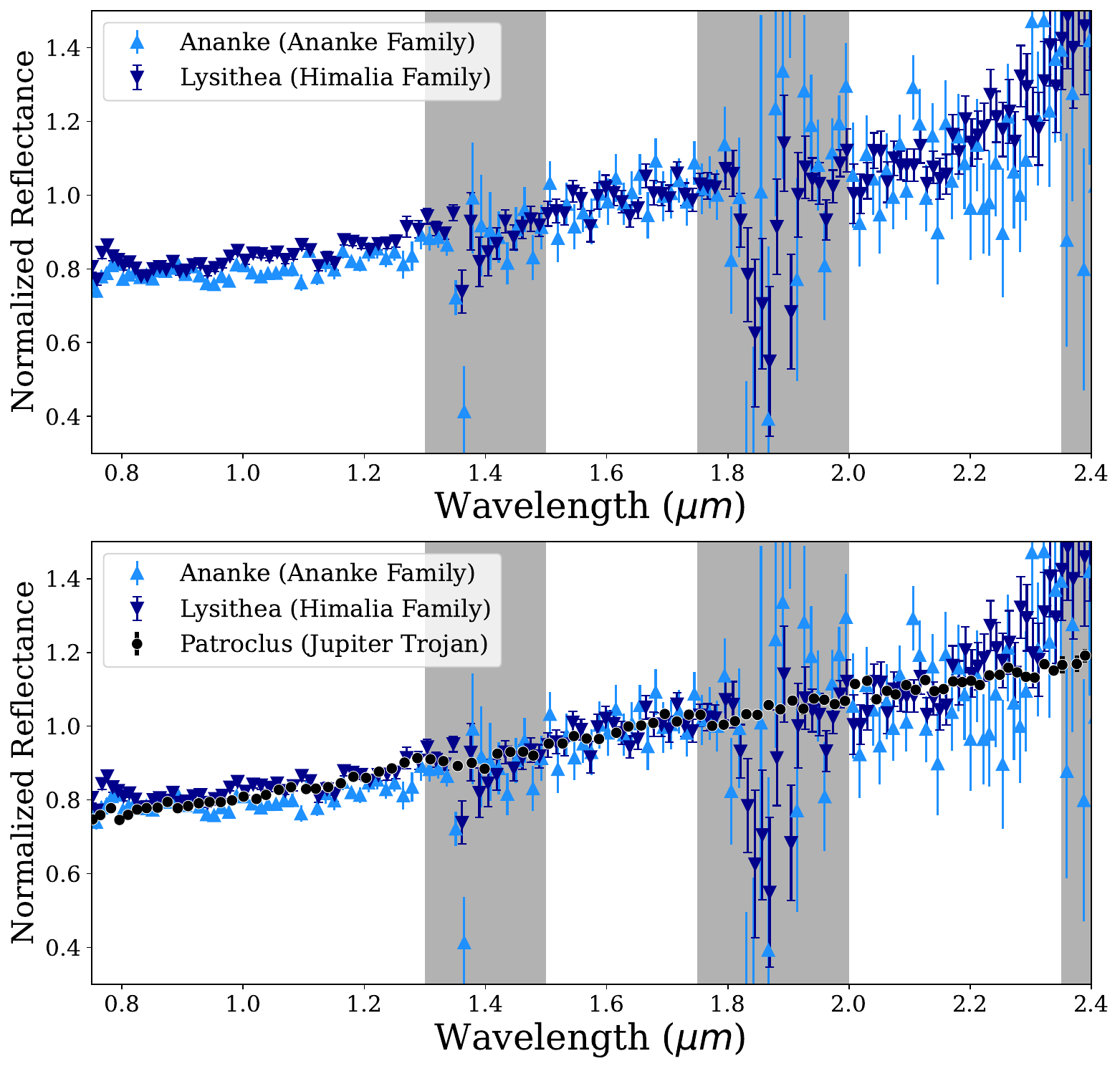}
\caption{Top: Spectra of Jovian irregular satellites Ananke and Lysithea. Bottom: Comparison to the Jovian Trojan and Lucy spacecraft mission target (617) Patroclus collected in \citet{Sharkey2019}. Lysithea is in the prograde Himalia family, while Ananke and its associated family is in a retrograde configuration. Both objects compare closely to the linearly red sloped P type spectrum of Patroclus. The gray bars represent wavelength regions with significant telluric absorption features that cause increased scatter and sometimes introduce spectral artifacts in observations of faint objects.}
\label{fig:c5_fig2}
\end{figure}

Motivated by the hypothesis that the Jovian irregular satellites represent material like the Jovian Trojans, we present the NIR spectra of irregular satellites Lysithea and Ananke in comparison to the Trojan (617) Patroclus, observed by \citet{Sharkey2019}. This comparison is presented in Figure \ref{fig:c5_fig2}, and we note strong similarity between the spectral behavior of all three of these objects. These spectra display neutral reflectances from $\sim0.8-1.0 \mu m$, transitioning to a linear, featureless red slope from $\sim 1.1-2.3\ \mu m$. We note that Lysithea is a prograde satellite and member of the Himalia family, while Ananke is the largest member of its own retrograde family.

\begin{figure}[h!]
\includegraphics[width=\textwidth]{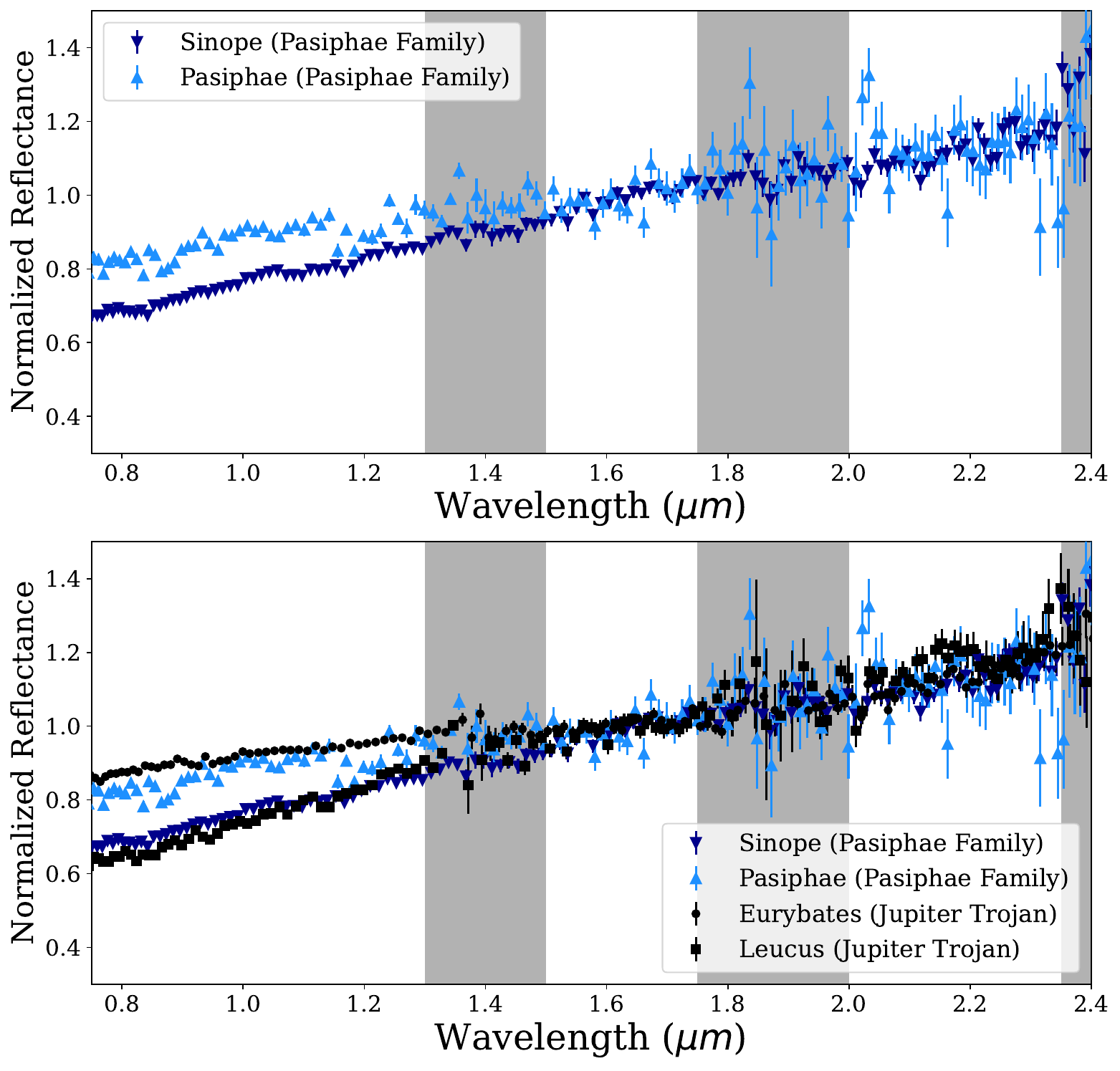}
\caption{Top: Comparison of Jovian irregular satellites Sinope and Pasiphae, which are both members of the retrograde Pasiphae family. Bottom: Sinope and Pasiphae present spectral differences which compare very similarly to the differences between Jovian Trojans (3548) Eurybates and (11351) Leucus collected in \citet{Sharkey2019}, which will be studied in detail by the Lucy spacecraft mission. The gray bars represent wavelength regions with significant telluric absorption features that cause increased scatter and sometimes introduce spectral artifacts in observations of faint objects.}
\label{fig:c5_fig3}
\end{figure}

Furthering the comparison between Jovian irregular satellites and Trojans, Figure \ref{fig:c5_fig3} presents a comparison between the newly acquired spectra of the Jovian irregular satellites Sinope and Pasiphae with previously collected observations of the Trojans (3548) Eurybates and (11351) Leucus \citep{Sharkey2019}. Pasiphae displays a slope from $0.8-2.4 \mu m$ similar to Ananke and Lysithea, but it is distinguished from those objects by showing neutral reflectance (i.e., slope near zero) from $1.4-1.7 \mu m$. Sinope displays a steeply red spectrum from $0.8-1.6 \mu m$, which rolls over to a less steep slope. Pasiphae and Sinope display spectral slopes similar to each other, as well as to Ananke and Lysithea, from $1.6-2.4 \mu m$.

The spectral behavior of (3548) Eurybates was found to be closely similar to Pasiphae, particularly in showing neutral reflectance from $1.4-1.7 \mu m$. (3548) Eurybates itself is notable as it has been identified as the largest member of a family of neutral, C-type-like objects, which are rare amongst the Trojan population \citep{Fornasier2007}. (11351) Leucus was found by \citet{Sharkey2019} to be representative of the redder class of Trojans identified by \citet{emery2011}. We note agreement in the spectral slope of Pasiphae between our observations and those of \citet{mebrown2000}, which shows an approximate rise of 10\% in reflectance over the range of 1.4-2.4 microns.

Overall, we find that the near-infrared spectra of Ananke and Lysithea compare closely to one another, and in turn, compare closely to the P-type Trojan (617) Patroclus. \citet{Grav2004a} report possible broadband evidence for water ice features in the NIR on Ananke, which we do not detect. Pasiphae compares closely to the more spectrally neutral Trojan (3548) Eurybates. Sinope compares closely to the D-type Trojan (11351) Leucus. These objects span a range of dynamical properties and include both prograde and retrograde orbits. A summary of the spectral properties for Jovian irregular satellites we observe, including their Jovian Trojan analogs, is given by Table \ref{tab:irregs_trojans_comp}. 

The visible spectra of Themisto and Leda are in general agreement with visible photometric measurements. Using combined visible and near-infrared photometry, \citet{Grav2004a} classify Leda, Lysithea, and Pasiphae as C-types, Sinope as a D-types, and Ananke and Themisto as P-types. We find our observations support these taxonomic assignments in all cases except for Lysithea. Instead, we find the near-infrared spectrum of Lysithea to track closely with Ananke (P-type). 

We find Lysithea's NIR slope to be more neutral (less red) than the photometric observations reported by \citet{Grav2004a}, with the discrepancy relating largely to the H-K slope (over the range $\sim1.6-2.2 \mu m$). We note that \citet{Grav2004a} report their spectrophotometry for Lysithea assuming V-J colors based on those measured for Himalia. The visible photometry shows both Lysithea and Ananke to have similar red slopes (data originally reported in \citealt{Grav2003}, discussed in the context of NIR observations by \citealt{Grav2004a}). Notably, the spectral shape that we report is more consistent with the updated classification of C/P type given by \citet{Grav2015}, which finds that Lysithea brightens from a visible albedo of $3.6 \pm 0.6 \%$ to an albedo of $6.9 \pm 1.1 \%$ at $3.4 \mu m$ (although the authors note that lightcurve sampling effects may be present). 

Near-infrared taxonomies give useful descriptions of spectral behavior, but without diagnostic absorption features they do not provide enough information to describe the mineralogical relationships between objects in different spectral classes. We provide further interpretations of possible relationships between P- and C-type spectra among the Jovian satellite population in the discussion section. 

\subsection{Uranian and Neptunian Systems}
\subsubsection{Uranian Satellites}

\begin{figure}[h!]
\includegraphics[width=0.5\textwidth]{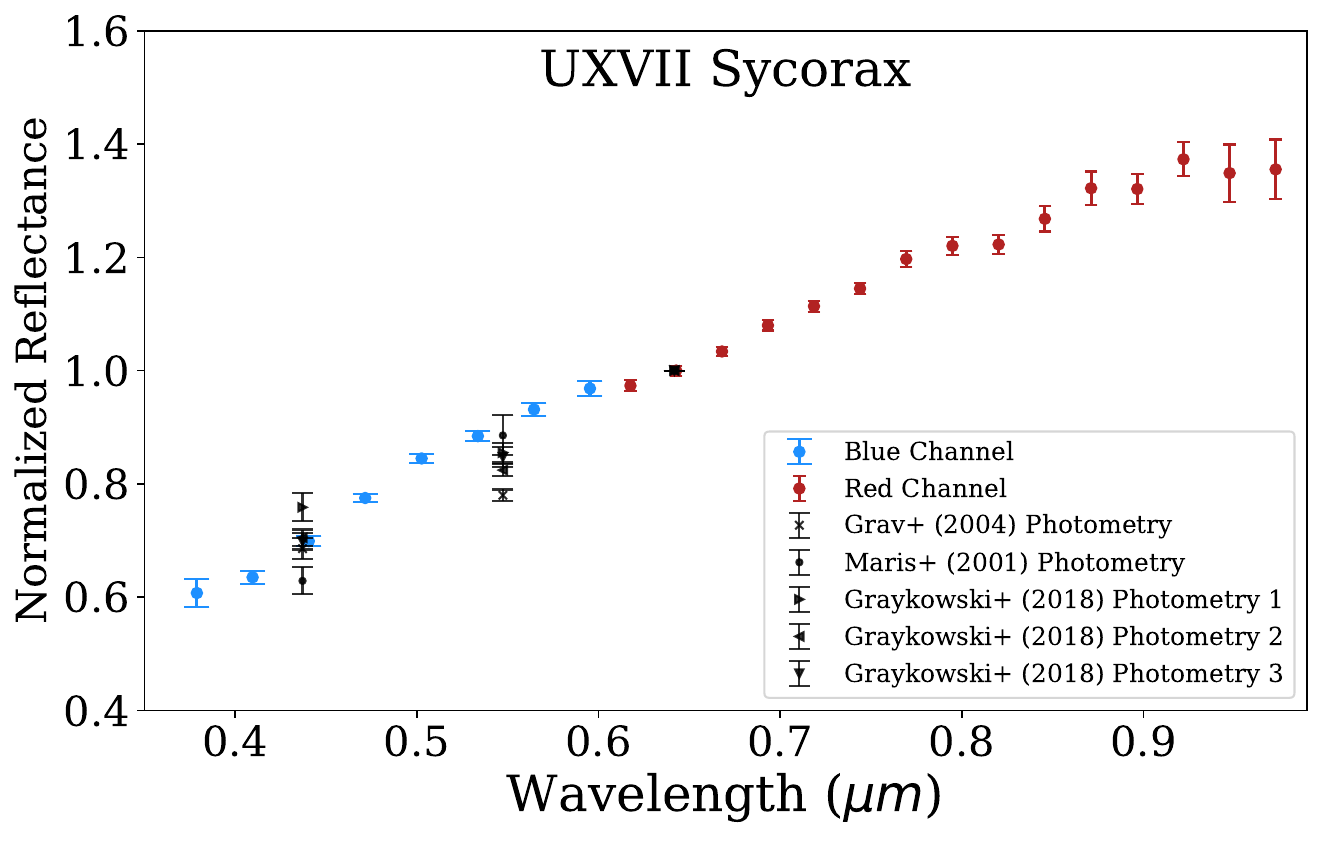}
\includegraphics[width=0.5\textwidth]{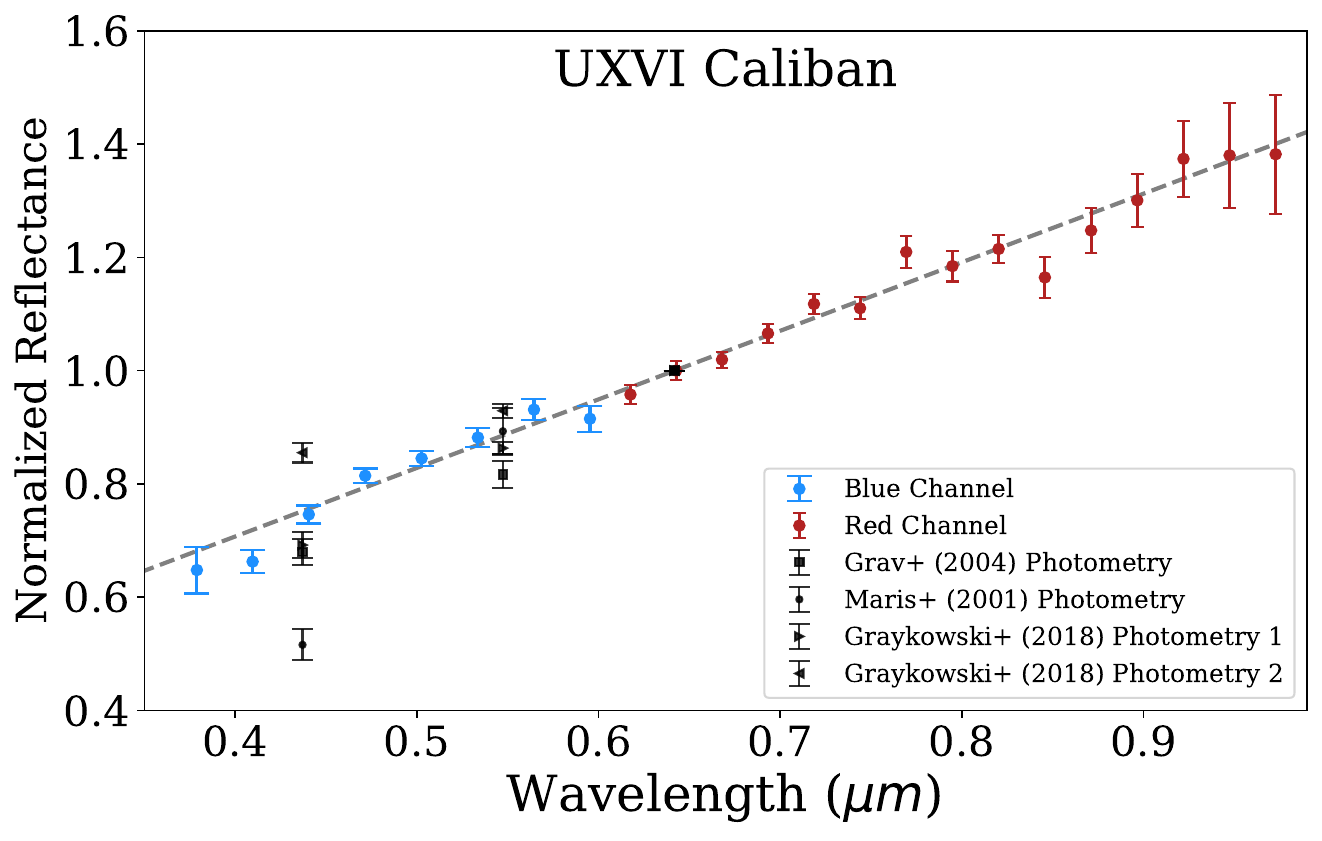}
\includegraphics[width=0.5\textwidth]{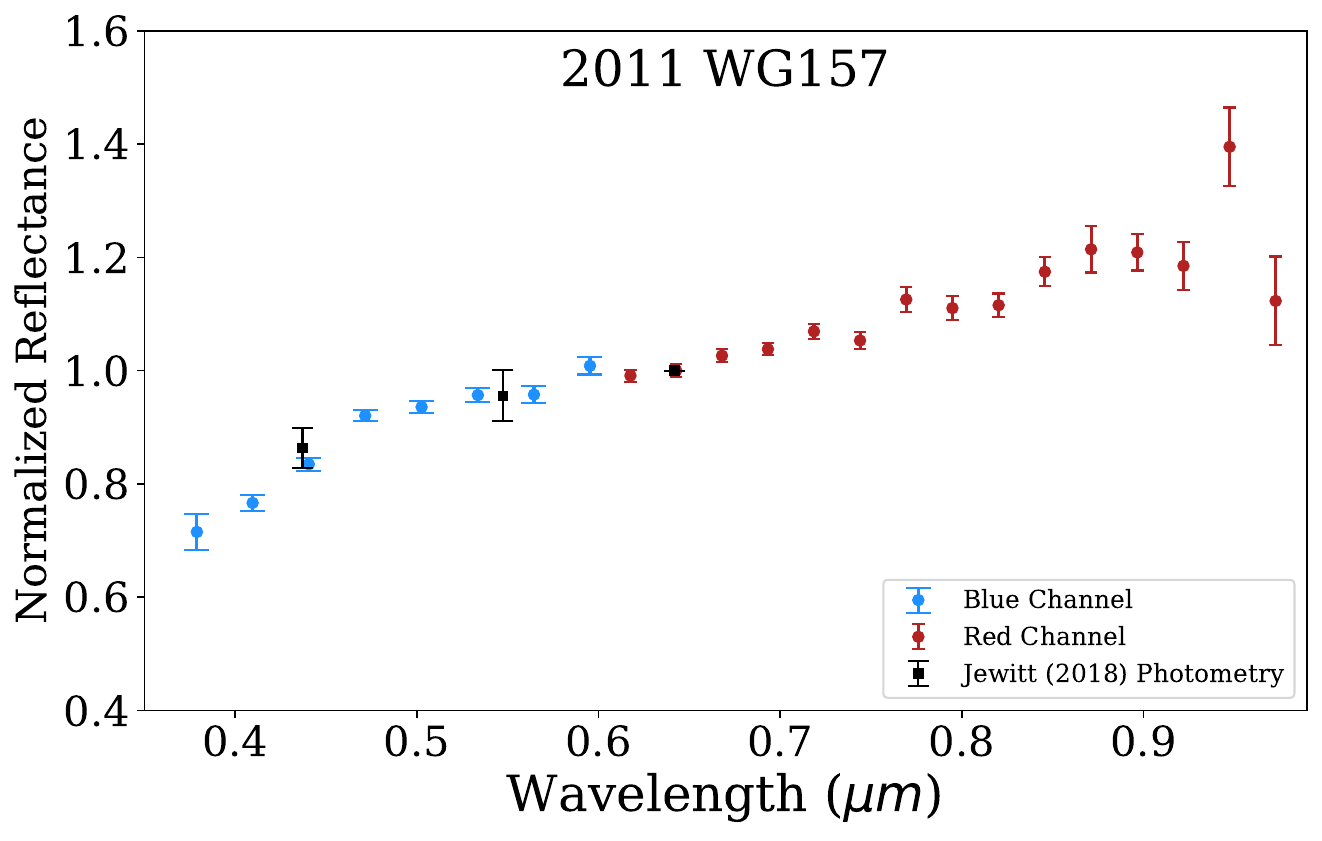}
\includegraphics[width=0.5\textwidth]{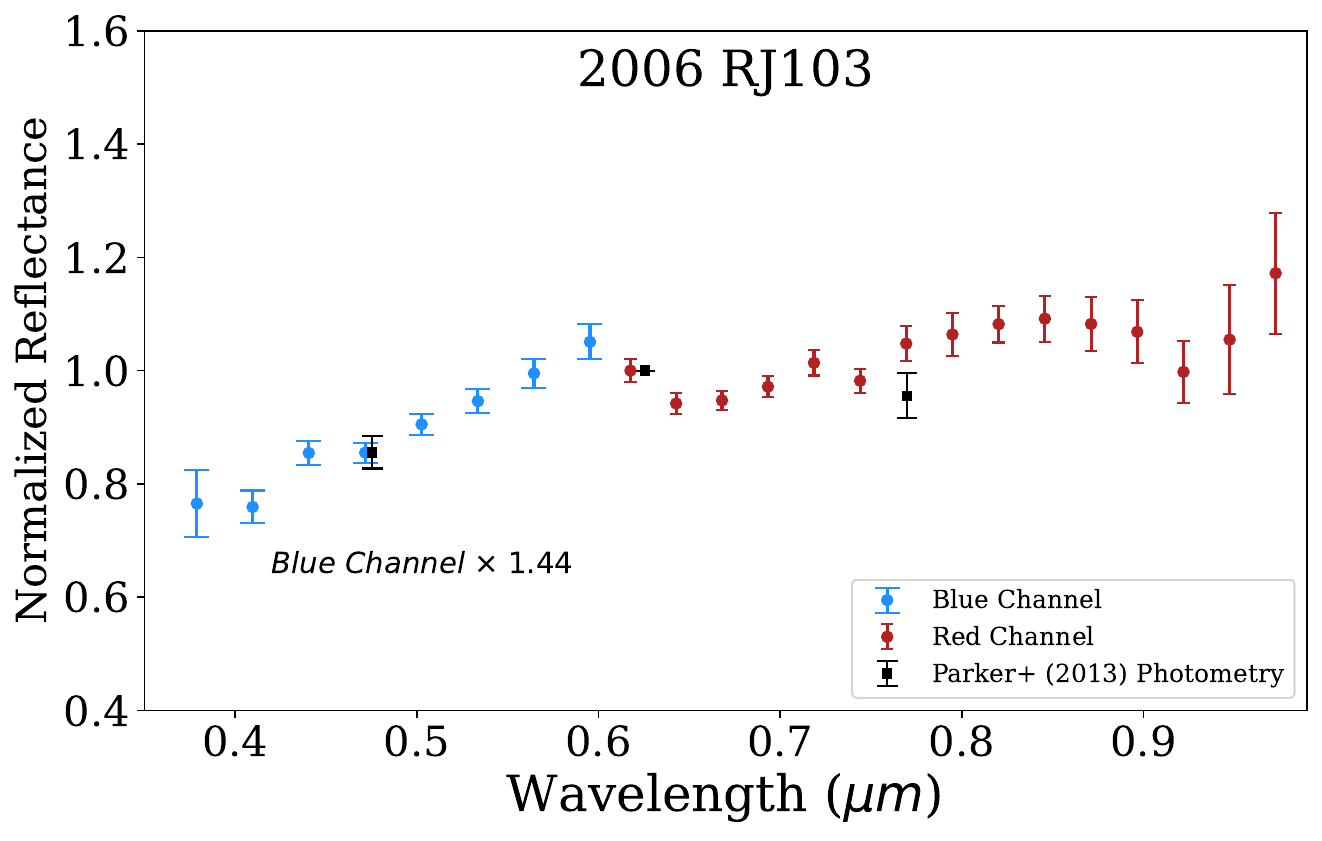}
\includegraphics[width=0.5\textwidth]{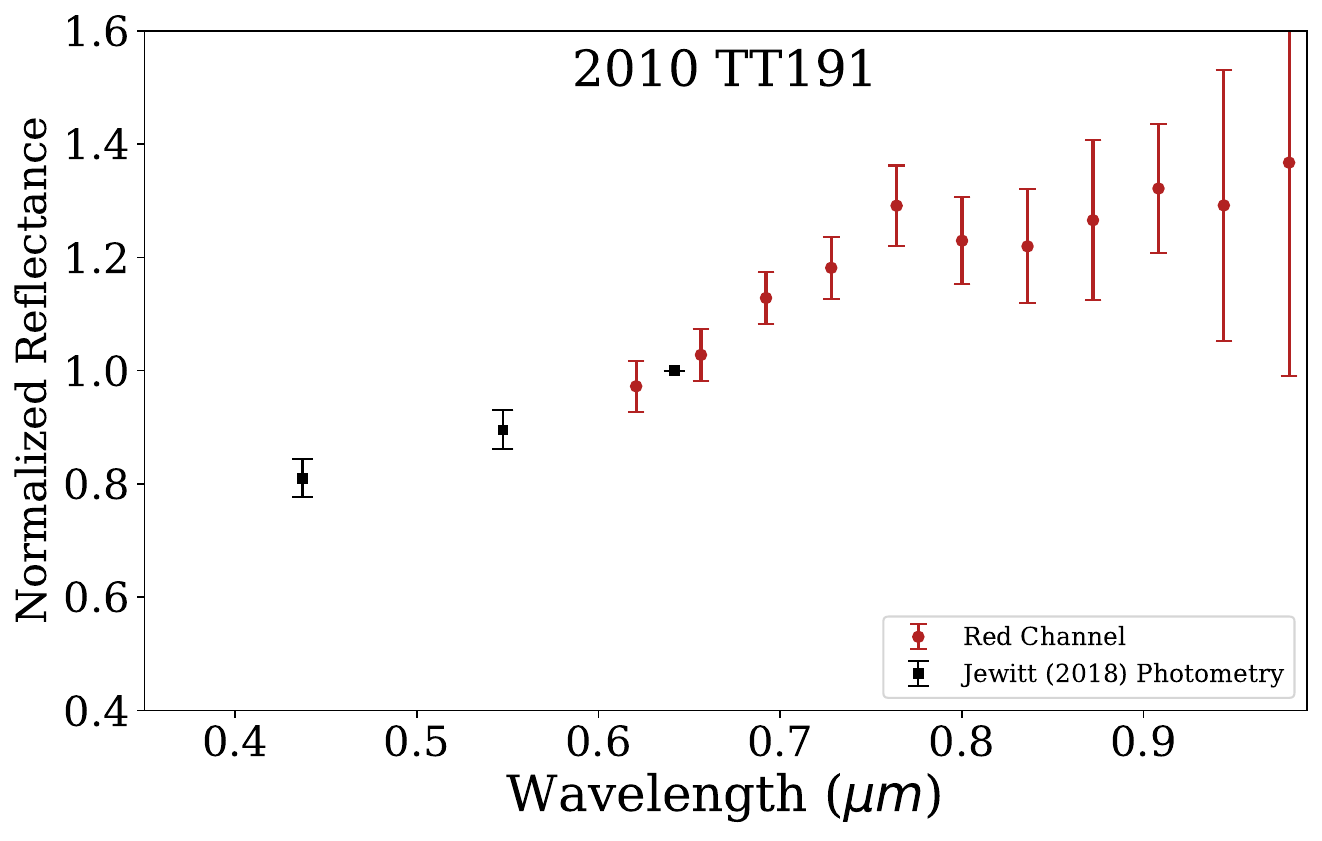}
\includegraphics[width=0.5\textwidth]{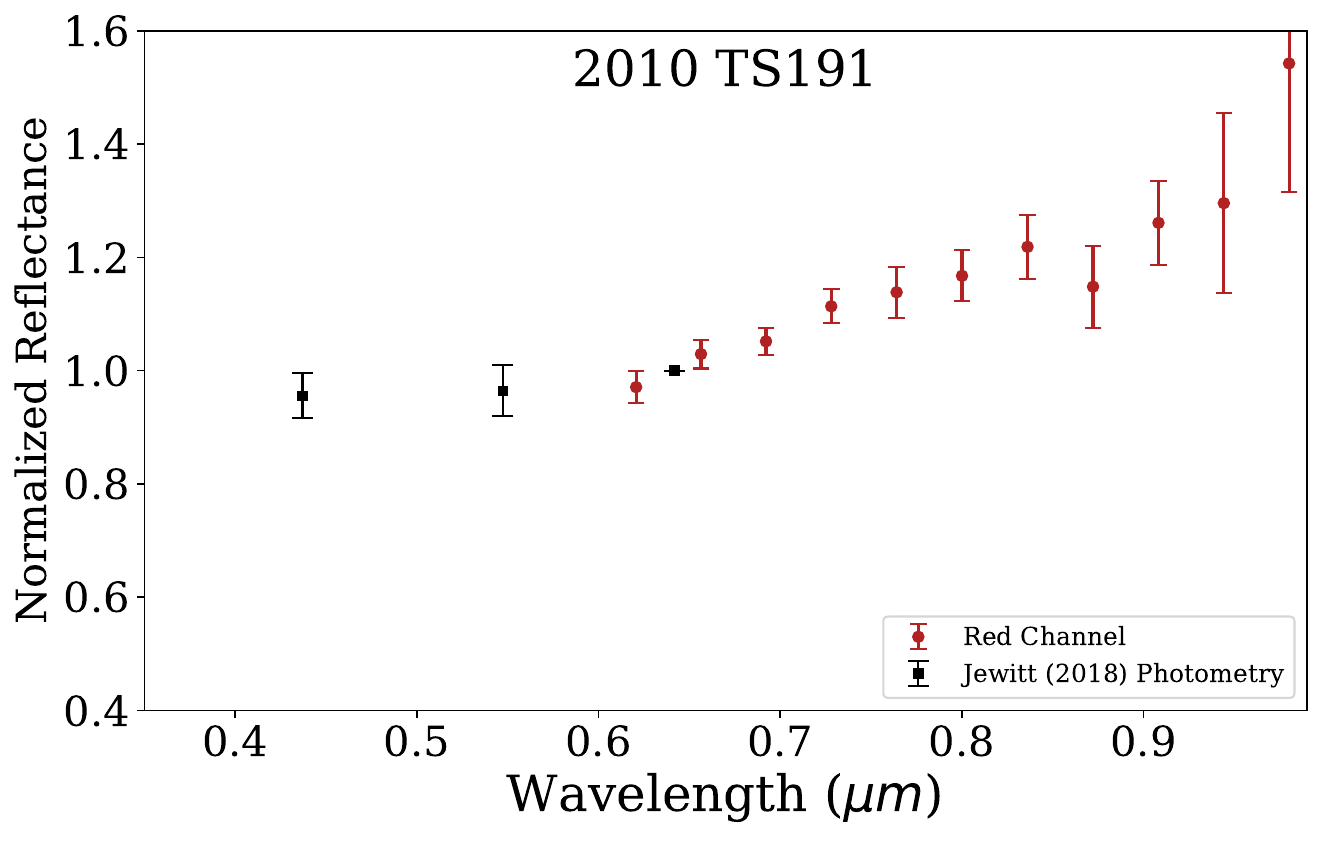}
\caption{Composite of spectra (channels: MODS1 Red and MODS2 Blue) collected of Uranian irregular satellites (Sycorax, Caliban) and Neptune Trojans (remaining panels) shown with comparison to photometric colors measured in the literature. Observations of 2010 TT191 and 2010 TS191 were not fully executed due to time constraints and did not produce recoverable traces in the blue channel. Except for the case of 2006 RJ103, the relative fluxes from the red and blue instruments are in good agreement and no arbitrary scaling is required.}
\label{fig:c5_fig4}
\end{figure}

The visible spectra of Sycorax and Caliban are presented in Figure \ref{fig:c5_fig4}. These two Uranian satellites display similarly steep red spectra. Since Uranus lacks a stable population of Trojans, we instead compare these objects to the colors of TNOs and Centaurs. Under this comparison, Sycorax and Caliban have visible spectra similar to the IR taxon \citep{Barucci2005,Perna2010}.  Objects in this color range are intermediate between the two modes of the Centaur color distribution, of which the majority of objects are clustered to grayer or redder colors (although there is significant variation, particularly amongst redder objects; see review by \citealt{Peixinho2020}). Spectral slope calculations are given in Table \ref{tab:icegiant_obj_table}.

Caliban displays a more linearly sloped spectrum without apparent concavities. While there is significant spread in the visible photometric colors we retrieve from the literature \citep{Maris2001,Grav2004b,Graykowski2018}, we note strong agreement with the mean colors from these combined datasets. We do not note any absorption features for the fainter satellite Caliban. Two inflections are apparent in Caliban's spectrum near $0.75 \mu m$ and $0.85 \mu m$, but they are both only one bin wide and we therefore conclude that our observations cannot resolve nor identify them as candidate features.

\begin{figure}[h!]
\includegraphics[width=.95\textwidth]{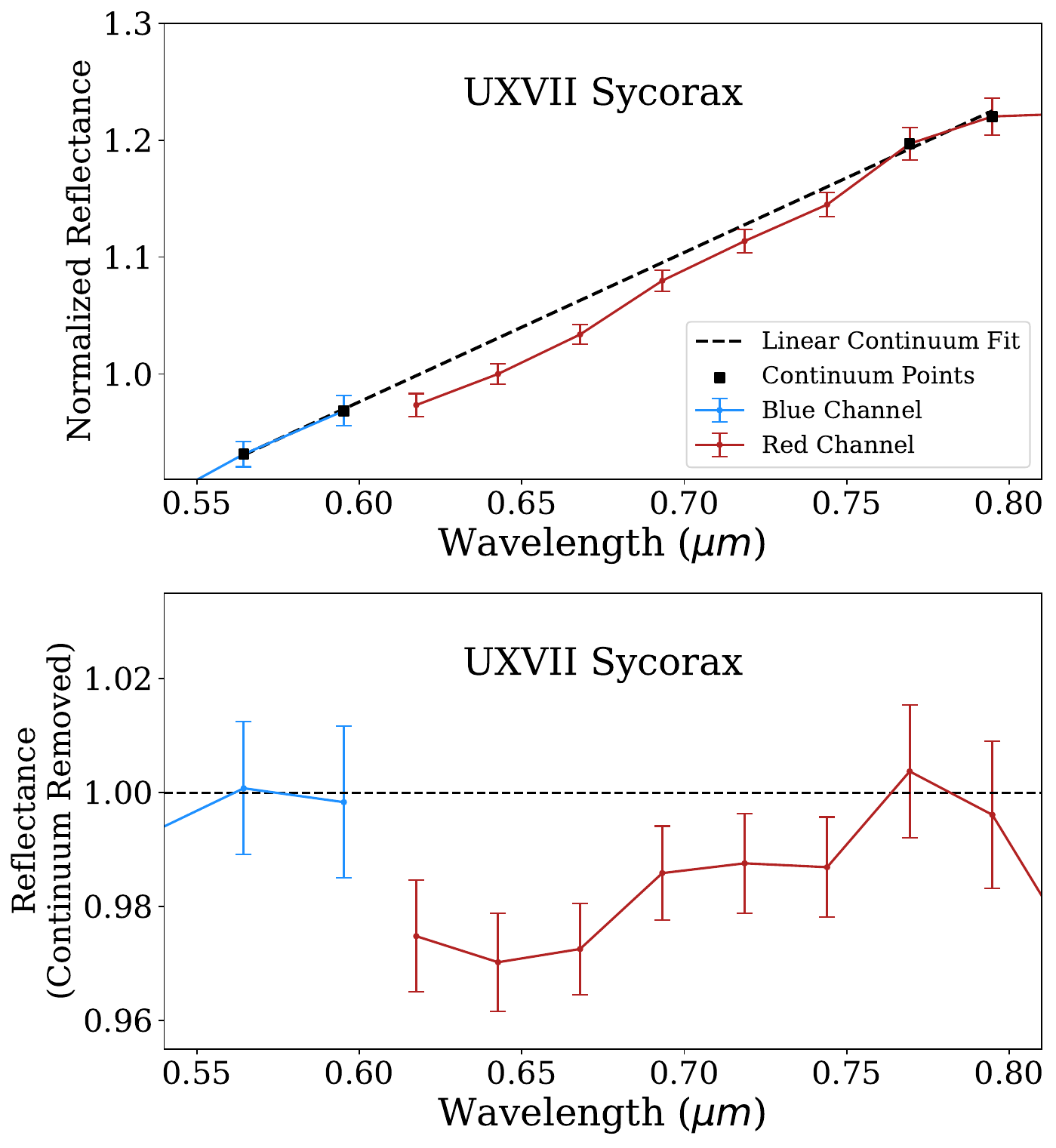}
\caption{\textbf{Top:} Spectrum of Uranian satellite Sycorax with a linear continuum fit as shown. A shallow dip is noted relative to this continuum from $\sim 0.65-0.75 \mu m$. This feature is not dominated by a single wavelength bin (which can be indicative of a noise spike or incomplete sky subtraction) and persists far from the edge of the effective wavelength range of the red instrument. Further assessment of this spectral region would be aided by higher SNR observations to constrain the continuum. \textbf{Bottom:} Continuum removed reflectance spectrum for Uranian satellite Sycorax. Here, values of unity indicate perfect agreement with extrapolation from the linear continuum. Variations below the continuum are small but statistically significant, consistent with the presence of an absorption feature. The depth of this feature is highly dependant on the choice of anchor points used in drawing the continuum, particularly along the short wavelength edge. Depths vary from $ 3.0 \pm 0.9 \%$ (using the continuum shown here) to $1.6 \pm 0.9 \% $ if all points between 0.55-0.65 $\mu m$ are included as anchors. }
\label{fig:c5_fig5}
\end{figure}

Further investigation of the $0.7 \mu m$ region of Sycorax's spectrum is given in Figure \ref{fig:c5_fig5}, which includes an example linear continuum drawn across either side of a possible absorption feature. Figure \ref{fig:c5_fig5} also displays the results of dividing the $0.5-0.8 \mu m$ wavelength region by the linear continuum shown. Sycorax displays a possible absorption feature near $0.7 \mu m$, with a band depth of $3.0 \pm 0.9 \%$ (where the depth as simply defined as the minimum value in the continuum removed spectrum).  This subtle feature requires careful choice of the continuum, particularly at the shorter wavelength edge between $0.55-0.65 \mu m$. For example, incorporating the red channel points between $0.60-0.65 \mu m$ while keeping the rightmost continuum defined between $0.75-0.80 \mu m$ still finds a significant feature between $0.65-0.75 \mu m$, but significantly decreases the depth to $1.6 \pm 0.9 \%$.  

While shallow, this feature is significant in comparison with the per-bin noise we achieved. We note that this feature is not near the edges of the effective wavelength of the red channel and is consistent across a wide wavelength range (it is not the result of an individual wavelength bin, which could indicate a noise spike or incomplete sky subtraction). This broad feature is inconsistent with systematic errors introduced by tracing and extracting faint sources. Such errors are typically monotonic as a function of wavelength and result in distortions to the continuum slope.  Absorption features near $0.7 \mu m$ are associated with $Fe^{2+}-Fe^{3+}$ charge transfer features in hydrated silicate materials, such as chlorites and serpentines, and is therefore interpreted as a marker of aqueous alteration in carbonaceous asteroids and meteorites \citep[e.g.,][]{King1989,Vilas1989,Vilas1994,Cloutis2011a,King1997}. 

\subsubsection{Neptune Trojans}

Figure \ref{fig:c5_fig4} also displays the reflectance spectra derived for four Neptune Trojans. In comparison to the Uranian satellites, the Neptune Trojans display more neutral visible spectra. We detect a significant drop in the reflectance of 2006 RJ103 near $0.70 \mu m$. While caution is warranted in interpreting shallow features in low SNR data, we note that, similar to our discussion of the feature in Sycorax's spectrum, this is robust across multiple wavelength bins and is not near the edge of the red channel. Wavelength-dependent slit losses may be present, as our spectrum differs from the colors reported from archival CFHT images by \citet{Parker2013} by $\sim 3 \sigma$. It is also possible that the rotational lightcurve of 2006 RJ103 provides an additional source of systematic error in the color measurements that must be corrected to compare to our spectrum. Regardless, such smoothly varying wavelength-dependent errors would not create spurious absorption features over a smaller portion of this wavelength range. This region is also outside of any area of strong, narrow atmospheric transmission or absorption features which could degrade the efficacy of sky subtraction. We further note that none of the other targets display this feature, and as such we conclude that this feature is significant and not likely to be a spurious result from data reduction. Although this feature is qualitatively different than the feature noted on Sycorax, its implications would be broadly the same: evidence for hydrated silicates indicative of aqueous alteration of surface material.

Our spectral observations of Neptune Trojan 2011 WG157 are in close agreement with the photometric colors reported by \citet{Jewitt2018}. At the blue end of our spectrum, we find a slope break near $0.45 \mu m$, but otherwise find an approximately linear spectrum from $0.5-0.9 \mu m$. Observations of 2010 TT191 and 2010 TS191, which were interrupted due to time constraints, only produced recoverable traces in the red channel and limit the comparison to literature photometry. Though these spectra are significantly noisier, we find that 2010 TT191 has a red slope from $0.6-0.9 \mu m$ which is consistent with a linear extrapolation of the colors reported by \citet{Jewitt2018}. We find that 2010 TS191 has a slope from $0.6-0.9 \mu m$ which is similar to 2011 WG157. Though 2010 TS191 and 2011 WG157 share similarities in the $0.6-0.9 \mu m$ range, 2011 WG157 appears redder from $0.4-0.6 \mu m$ in both the spectral and color data than the colors reported in this range by \citet{Jewitt2018}.

\begin{figure}[h!]
\includegraphics[width=\textwidth]{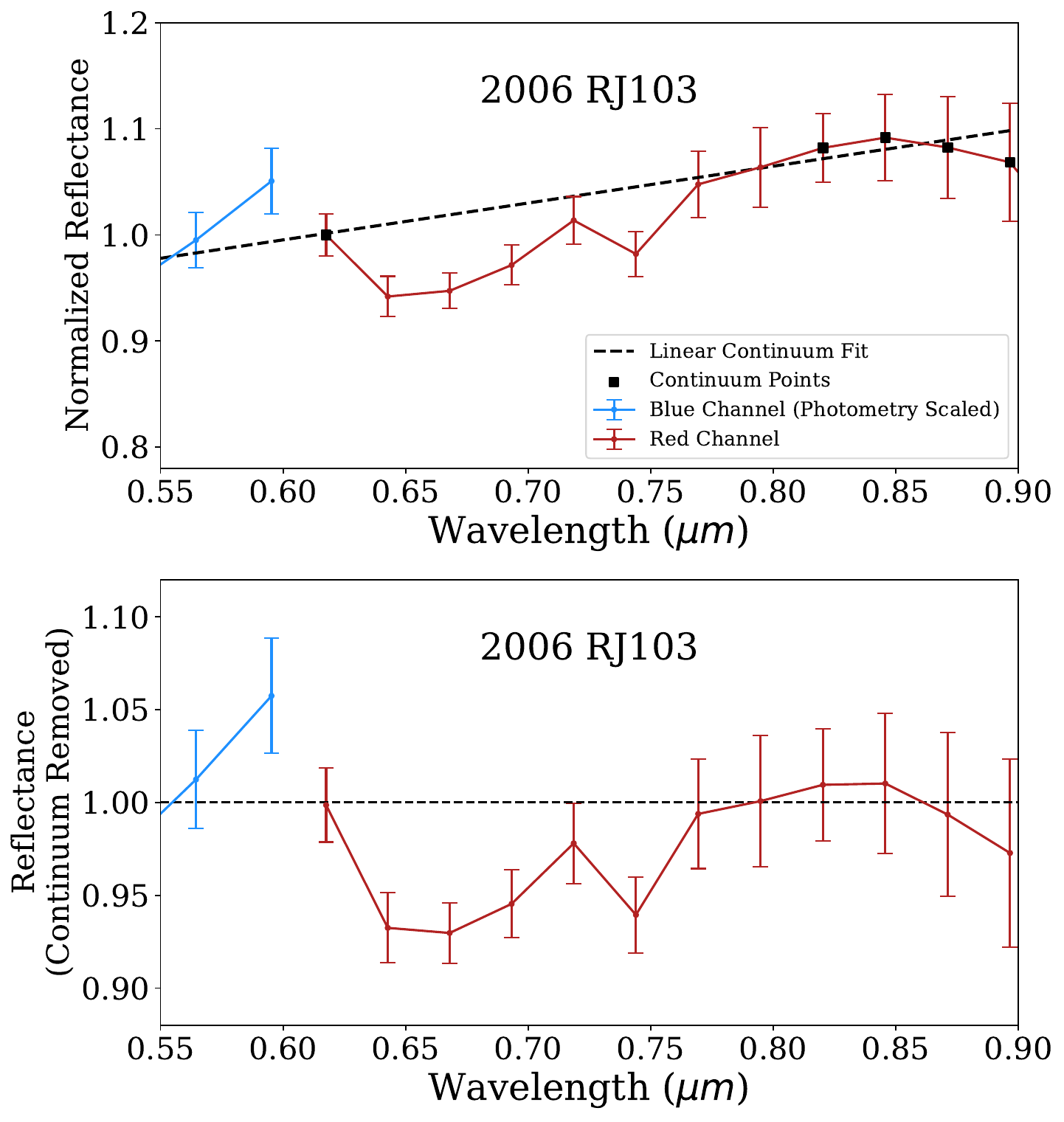}
\caption{\textbf{Top:} Spectrum of Neptune Trojan 2006 RJ103, with a linear fit to the spectral continuum along either side of the drop in reflectance detected near 0.7 microns. Each of a sequence of five wavelength bins appear inconsistent with this continuum at the $\sim 1-2 \sigma$ level. As with our discussion of the spectrum of Sycorax, this feature is not dominated by a single wavelength bin, is not near an area with significant telluric contamination, nor near the edge of the effective range of the red channel. \textbf{Bottom:} Continuum removed reflectance spectrum for Neptune Trojan 2006 RJ103. We note a drop in reflectance from $\sim 0.65-0.75 \mu m$, with a minimum near $0.64 \mu m$. We find the detection to be robust over a wide wavelength range. Absorption features in this wavelength region are often associated with $Fe^{2+}$ charge transfer features in hydrated silicate materials.}
\label{fig:c5_fig6}
\end{figure}

Figure \ref{fig:c5_fig6} displays the spectrum of 2006 RJ103 with an example continuum fit between $\sim 0.6-0.8 \mu m$. Careful definition of spectral continua is required for detailed characterization of features, as choices between polynomial order, as well as wavelength ranges to fit over, can affect resulting estimates of depth and centers of bands \citep{Mitchell2020}. However, simply establishing the significance of a feature detection is less sensitive to such choices. For transparency and to maintain consistency with the investigation of Sycorax, we opted to define a linear spectral continuum. The result of dividing the spectrum of 2006 RJ103 by this continuum is also displayed in Figure \ref{fig:c5_fig6}, which shows that all wavelength bins from $\sim 0.65-0.75 \mu m$ are below the continuum by $\sim 1-3 \sigma$. We conclude that this absorption feature is robust, as it is not near the edge of the effective wavelength range of the red channel, nor near areas of significant contamination from our telluric correction process (which could cause spurious, narrow-wavelength features). The feature is deeper $(7.0 \pm 1.6 \%)$ than the subtle detection noted for Sycorax over a similar wavelength region.

\section{Discussion}

\subsection{Regolith Grain Size Effects on C-/P-/D-Type Material Analogs}

We find a range of neutral to reddish spectra amongst Jovian irregular satellites, spanning C, P, and D asteroid taxonomic types. Each object in our observed sample of irregular satellites are close spectral analogs to the Jovian Trojan targets of the \textit{Lucy} mission. We are motivated to expand the characterization of Jovian irregular satellites from broadband photometric colors to spectroscopic measurements to allow for more detailed comparisons with other small body populations and to specific material analogs. But understanding the surface properties of individual irregular satellites requires exploring the possible relationships between spectral types. 

Previous observations of Himalia and its affiliated family \citep{Nesvorny2004} have investigated links between C-type objects among the Jovian irregular satellites and C-type objects in the main asteroid belt. \citet{Brown2014} found that Himalia's spectrum near 3 microns displays an absorption feature diagnostic of hydrated material, closely matching that of the outer main belt asteroid (52) Europa \citep{Takir2012}. Such a match suggests that at least some portion of the Jovian irregulars represent objects captured from the same parent source as some hydrated main belt asteroids. \citet{Bhatt2017} found the spectral match between Himalia and hydrated main belt asteroid (52) Europa extends to near-infrared wavelengths ($0.7-2.5 \mu m$). This suggests that objects too faint to be reasonably observed at $3 \mu m$, but which display spectral similarities to Himalia, are likely to have compositions including hydrated silicates (although confirmation would require a detailed assessment of how the presence and/or strength of the $3 \mu m$ band varies with object size). \citet{Bhatt2017} interpret their spectrum of Himalia, as well as other large C-type-like irregular satellites Elara and Carme, as consistent with materials found in hydrated carbonaceous chondrites. They suggest the spectral differences between these three objects can be largely attributed to differences in the dark opaque material(s) on the satellites' surfaces.

Although understanding the specific mineralogies of  the redder P- and D-type objects is speculative due to their lack of diagnostic spectral features in the NIR, comparisons to hydrated CI and CM chondrites provide useful starting points. Without specific compositional knowledge of these objects, we treat their spectral differences as a combination of both compositional and non-compositional causes. To explore non-compositional causes of spectral variation, we are limited to a very narrow range of studies that provide detailed reflectance measurements of CI and CM samples as a function of grain size and packing conditions. We do not claim any unique spectral match between our targets and a given meteorite, but instead aim to establish the magnitude of textural effects that are relevant for spectra of dark, hydrated materials. 

\begin{figure}[h!]
\includegraphics[width=\textwidth]{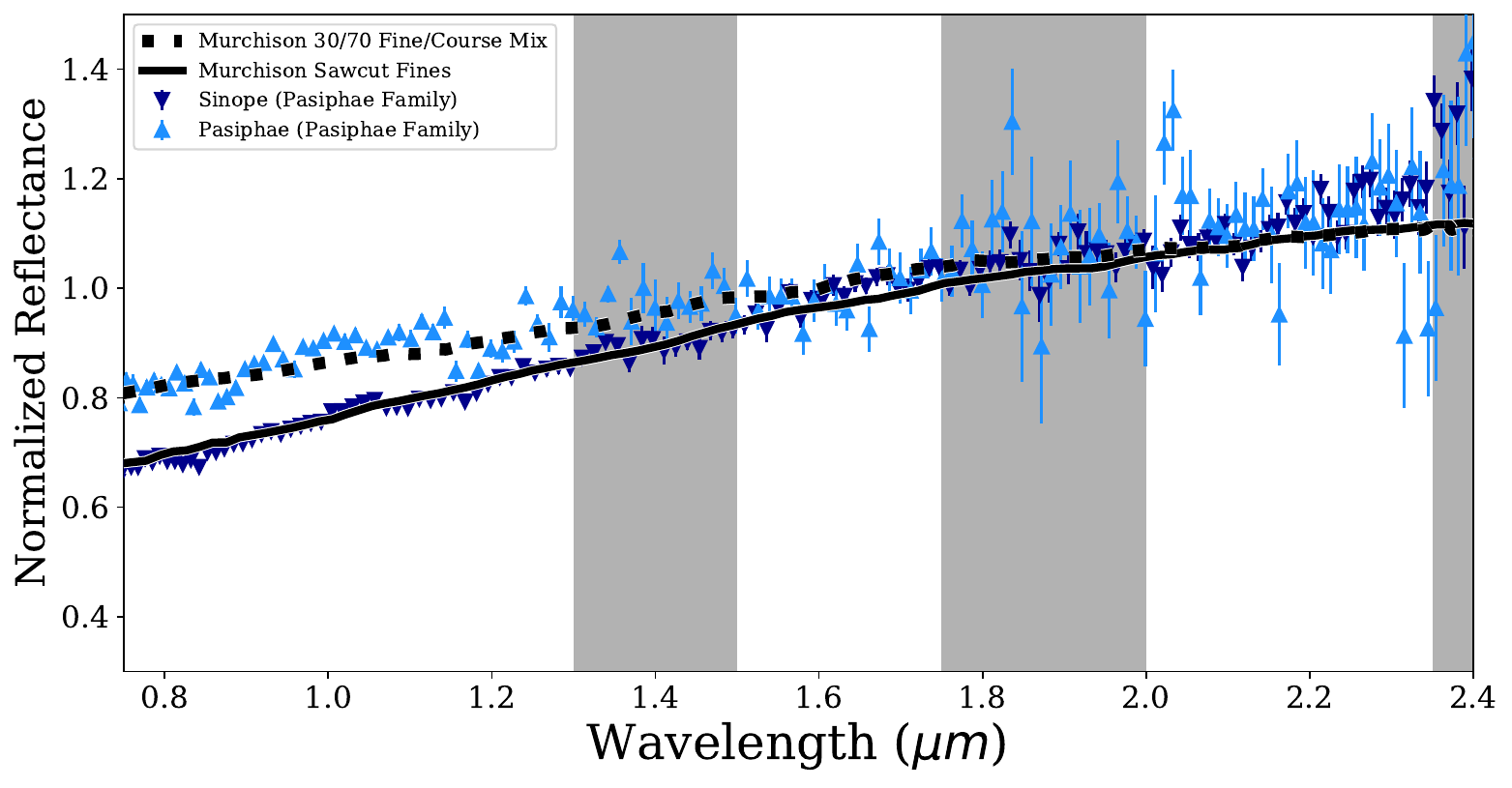}
\caption{Comparison of Jupiter irregular satellites Sinope and Pasiphae spectra with a laboratory spectra of carbonaceous chondrite meteorite Murchison samples prepared to different grain sizes, collected by \citet{Cloutis2018}, which were digitized to display for the convenience of the reader. We note that the scale of slope changes observed between these two members of the Pasiphae orbital family are consistent with slope changes associated with grain size variation in laboratory samples, and therefore it cannot be assumed that compositional effects are the only mechanism to explain spectral variation of these objects. The gray bars represent wavelength regions with significant telluric absorption features that cause increased scatter and sometimes introduce spectral artifacts in observations of faint objects.}
\label{fig:c5_fig7}
\end{figure}

We provide one example in Figure \ref{fig:c5_fig7}, displaying our observations of Sinope and Pasiphae with laboratory samples of the CM chondrite Murchison \citet{Cloutis2018}. The laboratory study of \citet{Cloutis2018} measured significant spectral slope and albedo differences as a function of grain size. They find the spectral slope (taken as the ratio of reflectance at $1.8 \mu m$ to that at $0.6 \mu m$) to vary from $\sim 1.05-1.60$ as samples varied from grains of size larger than $1000 \mu m$ to those smaller than $45 \mu m$. These variations in spectral slope are similar to the differences between the C-type (Pasiphae), P-type (Ananke, Lysithea), and D-type (Sinope) spectra we report in this work. While the spectral slope is sensitive to grain size, albedo variations are more subtle in this dark material, with albedo ranging from $\sim 4-6\%$. This range is comparable to both the average albedo ($4.0 \pm 0.8 \%$) and the full range ($2.9-5.7 \%$) of measured Jovian irregular satellite albedos measured in the 11-object NEOWISE sample \citep{Grav2015}. While this comparison between Murchison and the irregular satellites is not unique, the variety of spectra that Murchison presents as a function of grain size is notable. We conclude that if the irregular satellites are formed from similar material to the CM chondrites, grain size effects are critical to interpret their surfaces. This property is also important to account for when comparing the irregular satellites to other populations, especially those which have experienced differing collisional histories and may therefore have different grain sizes in their regoliths.

\subsection{Hypotheses for the Spectral Variation of Jovian Irregular Satellites and Trojans}

As discussed by \citet{Grav2015}, the spectral slopes and albedos of the Jovian irregular satellites compare similarly to Jovian Trojans and the Hilda asteroids. The spectrophotometric properties of Jovian irregular satellites have been noted to include neutral C-type objects, as well as moderately- to steeply-red P- to D-types \citep{Grav2004a,Rettig2001,Graykowski2018}. Thermal infrared observations from NEOWISE show that Jovian irregular satellites are quite dark, with albedos $\lesssim 0.05$ \citep{Grav2015}. The similarities between the albedos and colors of Jovian irregular satellites and Trojans already suggested that understanding the compositions of one population can greatly inform our understanding of the other. By finding close spectral matches between C-, D-, and P-type objects amongst both populations (Figs. \ref{fig:c5_fig2},\ref{fig:c5_fig3}), we emphasize the need for an origin scenario that explains both the variation within and the similarity between Jovian Trojans and irregular satellites.

We stress that it is difficult to assess the relative surface exposure ages between populations at different heliocentric distances via remote sensing alone. Strong heliocentric gradients in surface temperature, solar wind, cosmic ray bombardment, and dust particle bombardment all must be accounted for to compare populations in different locations within the outer solar system. Additionally, the spectral effects of irradiation on primitive, carbon-rich small bodies has been demonstrated to be complex and dependant on the compositional makeup of individual meteorite samples \citep[e.g.,][]{Lantz2018}, and the cumulative effects of solar wind alteration depends on a combination of competing processes that depend on the ions (He+ or H+) used in experiments \citep{Laczniak2021}. These complications highlight the usefulness of comparisons between populations within the same heliocentric region, as with Jovian irregular satellites and Jovian Trojans.

At present, the compositions of primitive objects near Jupiter are largely unconstrained. The Trojans have a well-established dichotomy in their visible and near infrared colors \citep[between ``red'' and ``less red'' groups, similar to D- and P-type asteroids, respectively,][]{Szabo2007,Roig2008,emery2011,Grav2012}. Spectroscopic observations at mid-infrared wavelengths find features that indicate the presence of porous silicates similar to cometary spectra \citep{Emery2006,Martin2023} with features that can be recreated via highly porous laboratory mixtures of olivine \citep{Martin2022} or with models of silicates embedded in a transparent matrix of salts \citep{Yang2013}. These findings inform efforts to model near-infrared Trojan spectra, which typically take the form of ``if\slash then'' hypothesis tests to identify plausible material analogs \citep[e.g.,][]{emery2004}. This approach has been used to explore possible physical causes of spectral differences amongst the \textit{Lucy} targets \citep{Sharkey2019}, which are featureless at visible \citep{SouzaFeliciano2020} and near-infrared \citep{Sharkey2019} wavelengths.

Sulfur-bearing compounds have been suggested as a cause for the reddened spectra seen amongst Jovian Trojans. Due to their spectral similarity, we argue that such a hypothesis also predicts the chemistries amongst Jovian irregular satellites, and perhaps similar but less processed forms of sulfur/ice chemistry amongst irregular satellites of the colder giant planets. \citet{Wong2016} suggests a link between Trojans and TNOs by hypothesizing Trojan surface evolution as tied to the presence or absence of $H_{2}S$. In support of this pathway, \citet{Mahjoub2016} demonstrated that electron irradiation and heating (from 50K to $>$120 K) of ice mixtures produces chemically distinct products depending on the presence or absence of $H_{2}S$. This includes radiolytic production of sulfur allotropes in cold (50K) samples that initially form as short, spectrally red sulfur chains that become longer, more spectrally neutral chains after heating \citep{Mahjoub2017}. The irradiated samples that included $H_{2}S$ showed strong absorptions in the UV, as well as absorption features at visible wavelengths that decrease in strength with additional heating to 200K \citep{Poston2018}. While these experiments demonstrate novel chemical pathways, it is noteworthy that the concentrations of volatile ices used (with abundances of methanol, ammonia, and $H_{2}S$ greater than that of water ice) are quite distinct from known concentrations seen in comets, which typically display much larger amounts of water \citep[with methonal and ammonia content typically $\sim1\%$ the abundance of water, e.g.,][]{DiSanti2017}. The effects that different volatile abundances may play in the overall evolution of irradiated and heated sulfur-bearing ices is unclear at present.

\citet{Cartwright2020} discuss how dust from collisions between irregular satellites could be a source for sulfur-bearing compounds on Callisto, which they preferentially detect on the Galilean satellite's leading hemisphere. A positive identification of a sulfur bearing species on a Jovian Trojan or irregular satellite would provide intriguing context for the overall origins of these objects. However, UV observations of Trojans by \citet{Wong2019} and \citet{Humes2022} do not detect any absorption features associated with $H_{2}S$, nor features related to the salts suggested by \citet{Yang2013}. We further note that our visible spectra of Sycorax and Caliban constrain the history of $H_{2}S$ in the ice giant region of the outer solar system, as their colder surfaces would more easily preserve the processed record of any sulfur-based chemistry similar to \citet{Poston2018}'s laboratory observations of absorption bands at 0.41, 0.62, and 0.90 microns. No such bands are observed in our spectra of these objects. We interpret the absence of these absorption bands to suggest that $H_{2}S$ is not a cause for the reddening observed on Uranian satellites. If $H_{2}S$ has played a role in these objects' chemical histories, then some other mechanism must have acted to remove any spectroscopic evidence.

\subsubsection{Incorporating Non-Compositional Effects}
Without a detected, direct compositional cause for spectral variation amongst Jovian Trojan and irregular satellites, textural effects are necessary to consider. There is no reason to \textit{a priori} assume that the irregular satellites and Trojans exhibit only textural differences or only compositional differences. It is, however, interesting to consider the hypothesis that textural distinctions act as the primary cause of spectral variation amongst these populations. Such a hypothesis for the Trojans must also explain why their colors are bimodal (though overlapping), implying that the two-color groups reflect populations with different regolith packing properties (i.e. grain size or porosity). Two distinct distributions of grain size or porosity could imply that the two groups of objects with similar compositions underwent distinct processing. Under this view, the red and less-red parent populations could have formed in similar regions of the protoplanetary disk but experienced different collisional histories.

However, this hypothesis is countered by the presence of red and less-red objects amongst the Jovian irregulars. This is because the irregular satellites are thought to be highly collisionally evolved, specifically exhibiting a size-frequency distribution consistent with fragmentation of a less-evolved, Jupiter Trojan-like population \citep{Bottke2010}. The presence of apparently collisionally evolved C, D, and P-type surfaces of varying sizes appears to disfavor collisional processing as the direct driver of spectral differences for the irregular satellites. 

The spectral variety of Jovian irregular satellites may also complicate the use of a primordial radiation crust as the main driver for the red/less-red color dichotomy amongst the Jovian Trojans. \citet{Nesvorny2018} and \citet{Marschall2022} find that the Trojan size-frequency distribution is likely primordial for objects larger than $\sim 10\ km$, and that the intrinsic impact probabilities amongst Trojans after their implantation are a factor of  $\sim 10^3$ smaller than those found  for the irregular satellites \citep{Bottke2010}. Therefore, in this size range, if the irregular satellites and the Trojans represent material captured from the same parent population(s), we would expect the irregulars to represent younger surfaces than similarly sized Trojans. As such, any initial radiation crust on an irregular satellite would be disrupted after their capture and would need to be subsequently reformed. Therefore, either lunar-style space weathering effects \citep[e.g., ][]{Gaffey2010} are not a significant cause for color distinctions among the irregular satellites, or space weathering occurs quickly and on material exposed on satellite surfaces after their capture. 

Alternatively, it is possible that if some irregular satellites contain volatile ices (or did so in their past), textural differences could be imparted by different activity histories. If one population does not contain enough volatiles to undergo periods of activity after a resurfacing event, but another population does, spectral differences could be introduced even if the non-volatile compositions of both populations are similar to each other. The hypothesis of activity-driven regolith changes may serve as an augmentation to the ideas discussed by \citet{Wong2016}, but critically it would mean that initial compositional differences in these populations would not directly manifest in the opaque components of their present-day regolith. Instead, it is perhaps possible that the presence or absence of primordial ices could manifest by producing different textural conditions (like regolith grain size or porosity) on their surfaces today.

\subsection{Uranian Irregulars and Neptune Trojans: Evidence for Hydrated Materials}

\citet{Graykowski2018} assess the irregular satellites of the four giant planets and suggest that they share similar color properties independent of heliocentric distance. We measure Sycorax and Caliban to have photometric B-R color ratios of $\sim 1.4$. This puts them intermediate between the two modes of the Centaur color distribution (compiled in \citealt{Peixinho2020}).

While statistics based on only two spectrally characterized objects is inherently speculative, the similarities of Uranus's two largest irregular satellites may suggest that Uranus captured its irregular satellite swarm from a more homogenous population than Jupiter's. The homogeneity of the two largest Uranian objects may alternatively suggest they experienced similar histories of activity/inactavity.

Notably, the spectral slopes of Sycorax and Caliban are consistent with laboratory spectra of the CI chondrite Alais \citep{Cloutis2011a}. As with our comparisons between Jovian irregular satellites and Murchison, we do not claim that these materials are directly analogous. However, Alais provides an interesting comparison of hydrated, CI material with reflectance properties that have been studied with variable grain size and texture with the same sample. Such studies of the textural and grain size properties on NIR reflectance of carbonaceous materials are scarce at present, and therefore we restrict our discussions to possible effects that current laboratory data can support. 

\begin{figure}[h!]
\includegraphics[width=.95\textwidth]{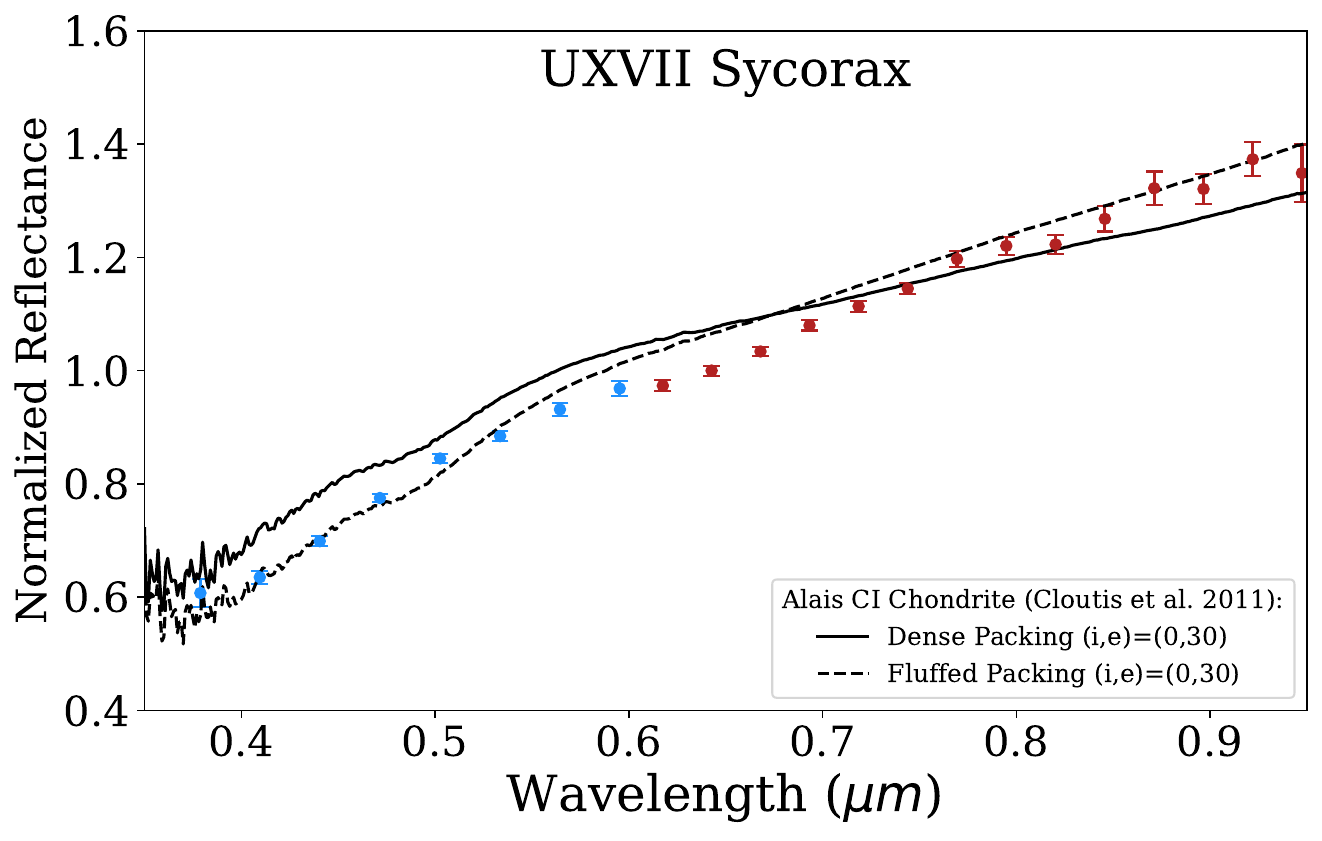}
\includegraphics[width=.95\textwidth]{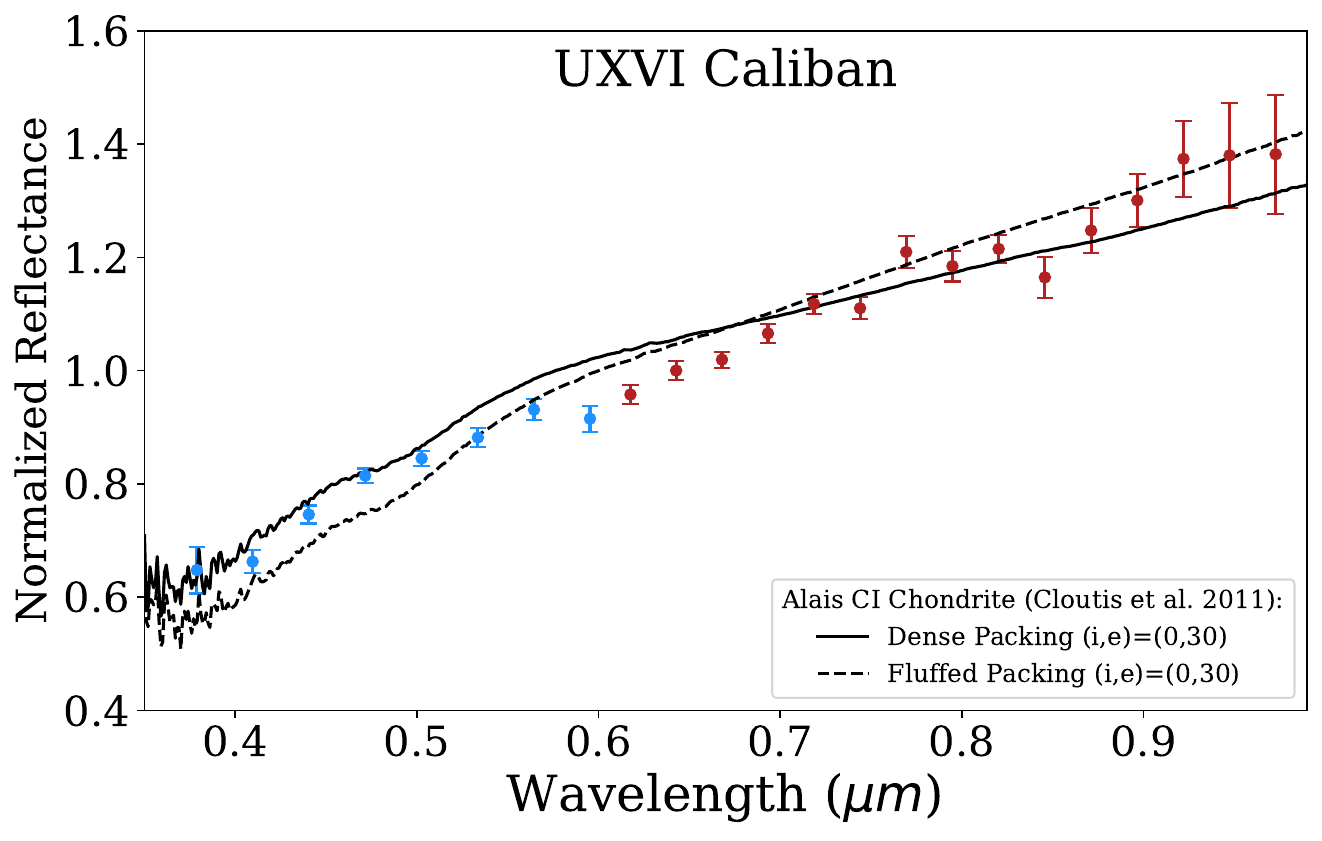}
\caption{Comparison of Uranian satellite Sycorax (top) and Caliban (bottom) with laboratory spectra of CI Chondrite Alais \citep{Cloutis2011a}. While spectral variation in the laboratory spectra was found due to textural, packing, and phase angle effects, the spectral similarities between Alais and Sycorax suggest that CI-like material may be a relevant analog for objects in the Uranian system. Spectra are scaled to have the same average value from 0.5-0.8 $\mu m$.}
\label{fig:c5_fig8}
\end{figure}

Figure \ref{fig:c5_fig8} presents a comparison between Sycorax and Caliban and the sample of meteorite Alais, and illustrates a plausible match in their visible colors. Recent study of the returned sample from asteroid Ryugu motivated a hypothesized link between CI-like material and material formed in the ice-giant region \citep{Hopp2022}. In this light, and with the evidence we present for a $0.7 \mu m$ feature on Sycorax, we conclude that the spectral similarities between Alais and Sycorax/Caliban strengthen the comparison between the irregular satellites of Uranus and hydrated carbonaceous chondrites. Additional studies of the spectral variability of CI chondrites may therefore provide useful insight into the variation of outer solar system spectra and enhance these comparisons. However, we lack diagnostic information linking these materials, and the match itself is inexact for both Sycorax (due to the absence of an apparant 0.7 micron feature) and Caliban (with some slope mismatch near 0.6-0.7 microns). We caution that significant spectral variability has been observed in this sample from non-compositional effects like packing, phase angle, and grain size \citep{Cloutis2011a}. These effects must be taken into account to provide direct matches between small bodies and dark, carbonaceous meteorites. This is also illustrated in Figure \ref{fig:c5_fig8}, which shows the spectral variations that texture/packing can introduce to the same sample of Alais.

We note that \citet{Romon2001} obtained (non-simultaneous) visible and near-infrared spectra of Sycorax. They found a sharp drop in reflectance from visible to NIR colors. This spectrum, red in the visible but blue in the infrared (dropping in reflectance by $\sim50\%$ from $0.80-1.25 \mu m$) defied explanation via their spectral models. This may be reconciled by invoking significant rotational variation on its surface. \citet{FarkasTakacs2017} find that Sycorax has a $\sim 6.9$-hour rotational period and an amplitude of 0.12 magnitudes, seemingly too little variation to explain this discrepancy. \citet{Paradis2019} found evidence for latitude-dependent albedo variations as high as 50\% on the inner moons of Uranus by comparing observations taken between the 1997 and 2015. If Sycorax displays similar albedo variations, it could indicate that this phenomenon is prevalent among other small Uranian satellites.

\citet{FarkasTakacs2017} find that Sycorax and Caliban have measurably different albedos, of $0.065_{-0.011}^{+0.015}$ and $0.22_{-0.12}^{+0.20}$, respectively. The lower range albedo for Sycorax is consistent with a carbonaceous surface, similar to D-type asteroids and Jovian Trojans \citep{Grav2012}. However, the brighter nominal albedo of Caliban would likely imply an ice-rich surface, similar to the five Uranian classical satellites \citep{karkoschka1997,Paradis2019} and to the Neptunian irregular satellite Nereid \citep{buratti1997,Kiss2016}. The large uncertainty on the albedo estimates of Caliban obscures further characterization, as its lower bound is also consistent with the brighter end of the Jovian Trojan distribution. The similarities in their optical spectral slopes suggest that their colors are dominated by similar reddened opaque materials. However, if a larger albedo difference exists, consistent with the nominal values of \citet{FarkasTakacs2017}, then the presence of an absorption band on the darker object Sycorax could imply that these two satellites have experienced different thermal histories. However, while the $0.7 \mu m$ feature has been well-linked with hydrated materials, other hydrated asteroids have been observed to lack this feature \citep{Howell2011}, so its absence is not itself diagnostic. 

\citet{Jewitt2018} discusses how the neutral color distribution of the Neptune Trojans appears consistent with the Jovian Trojans. However, further observations \citep{Markwardt2023b} suggest abundances of ultra-red Neptune Trojans at a rate more typical of TNOs than to Jupiter Trojans. A confirmed feature on 2006 RJ103 would further complicate a direct connection between the surfaces of Neptune and Jupiter Trojans, as the Jovian Trojans are featureless in the visible and near infrared. 

\citet{Seccull2018} find evidence for a $0.7 \mu m$ feature on the Kuiper belt object 2004 EW95. Confirmation of this candidate feature on 2006 RJ103, and further studies to characterize the spectral variability of Neptune Trojans will be critical to clarify the links between this population and to provide direct compositional assessments between them and other small body populations.

\subsection{Irregular Satellite Thermal Histories: Connections to Kuiper Belt Object Interiors?}

Given the possible origins of irregular satellites from the PKB, combined with intense collisional evolution after their capture as satellites, there is a possible connection between irregular satellites' surface compositions and the interiors of present-day KBOs. For example, if a large irregular satellite were pummeled by collisions, would some of those impact events dredge up materials from their deep interiors? If so, what story would they tell us about the nature of early heating within such bodies?

Recent models of KBO binary formation suggest most were created by the so-called streaming instability model, which states that millimeter/centimeter-sized particles (pebbles) can be aerodynamically collected into self-gravitating clouds that then directly collapse into planetesimals \citep{Nesvorny2020,Nesvorny2021}. It is suspected that this mechanism explains how most planetesimals in the PKB were formed, with the majority of binaries stripped during Neptune's migration across the PKB. The largest population of binaries found today is in the cold classical Kuiper belt, a region between 40 and $\sim47$ au that has low eccentricities and inclinations \citep{Noll2020}. The home of worlds like Arrokoth, objects from this region formed in situ and were never disturbed by Neptune.

Once created, PKB objects could have been heated by the decay of the short-lived radiogenic nuclide $^{26}$Al, which has a half-life of $\sim0.71$ My. Objects that formed early and/or were large enough might have experienced enough heating in their interiors to melt and produce water, which in turn could have led to aqueous alteration among the rocky materials residing in the same region. Conversely, objects taking many Myr to form would have not have been substantially heated, and so may retain their primordial nature (e.g., Arrokoth).  

The evidence for deep heating and aqueous alteration in 100-300 km objects from the PKB is currently a mixed bag. On the positive side, planetesimal formation in the PKB had to take place early enough to form the order of $\sim10^8$ objects prior to solar nebula dissipation. If these objects formed within the first few Myr of CAIs, as expected, they should have been heated by short-lived radioisotopes (e.g. Prialnik et al. 2008; Shchuko et al. 2014). As supporting evidence,  consider (87) Sylvia, a 280 km diameter P-type asteroid in the outer main belt.  Dynamical models indicate that Sylvia was likely captured from the destabilized population during the giant planet instability \citep{Vokrouhlicky2016}. While Sylvia's exterior is spectrally similar to anhydrous materials, studies of its gravitational interactions with its tiny satellites point to Sylvia having a differentiated interior \citep{Carry2021}. In fact, the thermal modeling work done by \citet{Carry2021} suggests Sylvia has a three-layer structure: a central region dominated by a muddy ocean, a porous layer free of water, and then a primordial outer layer that remains too cold for ice to melt. This structure is consistent with its bulk density, which is $1.378 \pm 0.045\ g/cm^3$. While it is currently unclear whether Sylvia is representative of all KBOs of similar size, it does raise the possibility that some large irregular satellites could have had aqueously altered materials at depth exposed by collisions.

On the more negative side, 5 of the 7 KBO and scattered disk binaries 100-300 km in diameter with measured bulk densities have values that are below 1 $g/cm^3$ \citep[][see Table \ref{tab:TNO_densities}]{Grundy2007,Stansberry2012,Vilenius2012,Vilenius2014,Bierson2019}. These low values indicate these KBOs have substantial porosities, and modeling work suggests that too much Al heating would have melted the objects enough to remove this porosity \citep{Bierson2019}. A potential way out of this predicament would be for most KBO binaries to form at least 5 Myr after CAIs \citep[e.g.,][]{Bierson2019}. This would remove $^{26}$Al heating as an issue, allowing the KBOs to preserve porosity produced by the planetesimal formation process. The potential issues with that scenario are that (i) it assumes a very long-lived solar nebula in the outer solar system, and (ii) that the vast majority of planetesimal formation would need to start fairly late after CAIs. We will argue, however, that there is a middle ground on this issue, and will return to it below.

While data is limited, irregular satellites appear to fall on the positive side of the ledger, in that they appear to have bulk densities that are comparable to (87) Sylvia. For example, Saturn's irregular satellite Phoebe has a bulk density of 1.67 $g/cm^3$, higher than most of the known KBO bulk densities \citep{Thomas2007}. Jupiter's moon Himalia was inferred to have a bulk density of 1.63 $g/cm^3$, assuming a mean radius of $\sim$85 km \citep{Cruikshank1977}, while numerical simulations of Himalia's perturbations on the orbits of other nearby irregulars suggest values between 1.55 and 2.26 $g/cm^3$ \citep{Brozovic2017}. Phoebe and Himalia also show evidence for aqueous altered materials and perhaps an ice-rich subsurface \citep{Brown2014,Clark2005,Fraser2018}.  If so, these bodies would fit in well with the idea that higher bulk density values are consistent with melting \citep{Castillo-Rogez2012}.

The results from above would seem to leave us in a quandary. Some KBOs and scattered disk binaries have low enough bulk densities that they had to form well after the decay of $^{26}$Al, while other KBOs, or objects derived from the destabilized population, appear to have high enough bulk densities that they experienced internal melting. A possible solution may come from an examination of the KBO and scattered disk binaries shown in Table \ref{tab:TNO_densities}. Of the five objects listed with bulk densities $<$ 1 $g/cm^3$, three appear to be residents of the cold classical Kuiper belt (shown with shaded backgrounds in Table \ref{tab:TNO_densities}).  Cold classical KBOs formed further from the Sun than any other KBO (i.e., 40-47 au vs. 20-30 au for most objects).  If any KBOs were to form late in the lifetime of the solar nebula, it would be these bodies.  It is likely that in the cold classical region, photoevaporation of the solar nebula was needed to decrease the gas to dust ratio to a value suitable for planetesimal formation via the streaming instability \citep{Nesvorny2021}. This effect may have delayed planetesimal formation to timescales occurring after $^{26}$Al depletion.

If we were to remove the cold classical objects from consideration, we find that there is a 50-50 chance that the remaining bodies have bulk densities $>$ 1 $g/cm^3$.  These bodies likely formed between 20-30 au in the PKB. Given their proximity to the giant planet zone, where the cores of the giant planets had to form prior to loss of the solar nebula, and C-type asteroids and carbonaceous chondrite meteorites likely formed within 2-3 Myr of CAIs, we postulate that planetesimal formation processes occurred earlier here than in the cold classical Kuiper belt. Earlier start times would have allowed some objects to experience interior melting via $^{26}$Al, while others perhaps formed too late to experience much melting. Accordingly, (87) Sylvia, Himalia, Phoebe, (65489) Ceto/Phorcys, and (612239) (2001 QC298) were arguably among the earlier objects to form in the PKB, while (42355) Typhon–Echidna and (47171) Lempo were not (Table \ref{tab:TNO_densities}).  Another key consideration is that $^{26}$Al heating would work from the inside out, so objects with melted interiors could still have primitive porous exteriors. Partial melting could hide the characteristics of these objects from view, as suggested by the anhydrous material on (87) Sylvia. 

Finally, collisional evolution may have also played some role in this story. For example, we find it plausible that impacts removed porous exterior materials from Phoebe and Himalia, leaving behind bodies denser than (87) Sylvia. Conversely, collisions may have also shattered some objects from the PKB, potentially leaving them with lower bulk densities. At this time, it is difficult to say where these putative effects come into play for different objects. But, what we can say is that disruptive collisions among 100-300 km diameter bodies may mix interior and exterior materials in the surviving parent body and in their fragments. This would mean that some aqueous altered materials could be left exposed for detection from remote observations and might be found in some comets that approach Earth.  This makes the irregular satellites rich population to explore, with the largest bodies likely to have been strongly affected by impacts in some manner.

\section{Conclusions}

We extend spectroscopic coverage of the Jovian irregular satellites to objects with $D<10s$ of km. Our Jovian irregular satellite spectra closely match the spectra of Trojan asteroids, including the \textit{Lucy} targets. Previous observations have established that the Jovian satellite Himalia is hydrated \citep{Brown2014} and similar to C-type asteroids \citep{Brown2014,Bhatt2017}, and we find the smaller Himalia family member Leda to have a spectrum consistent with a C-type. If Leda and the smaller objects in the family retain evidence for hydration, that would imply that collisional breakup of this family has not acted to erase this record. 

Samples of the CM chondrite meteorite Murchison provide reasonable near-infrared spectral analogs to our sample of observed Jovian irregular satellites and Trojans. We emphasize that samples of Murchison prepared to different grain sizes display spectral variation of a similar magnitude to our observed object-to-object spectral variations in these populations. We therefore suggest that textural differences must be accounted for when making compositional comparisons between these dark, possibly hydrated objects. We hypothesize that the presence or absence of volatile ices on irregular satellite or Trojan surfaces could also be a driver for spectral differences, as past activity could lead to present-day distinctions in regolith grain sizes.

We find evidence for a $0.7 \mu m$ feature on the Uranian satellite Sycorax. Recent analysis suggests that Ryugu's isotopic properties place its formation (as well as that of other CI chondrites) to an outer solar system reservoir near to the ice giants' formation regions in the protoplanetary disk \citep{Hopp2022}. While not uniquely supportive of a compositional match, the spectral similarities between Sycorax, Caliban, and the CI chondrite Alais shows that CI chondrites provide plausible matches to surfaces in the Uranian system. The wide spectral diversity of outer solar system materials can be aided by analogies to the wide variety of spectroscopic properties found among CI and CM chondrites. The widespread presence of hydrated material(s) in the present-day outer solar system would provide one observational link between CI samples and small bodies near the ice giants. 

Given the presence of hydrated material on Himalia \citep{Brown2014} and Phoebe \citep{Clark2005,Fraser2018}, and the evidence we provide for hydration on Sycorax in this work, it is possible that giant planet irregular satellites generally represent an additional source of hydrated material in the solar system. Our detection of a candidate absorption feature on Neptune Trojan 2006 RJ103 would raise the possibility that the Neptune Trojans also contain hydrated material and provide evidence that the Neptune Trojans are distinct from the Jovian Trojans despite their similar optical colors \citep{Jewitt2018}. Further spectral studies of these and other TNOs can constrain their material properties and give insight into the alteration histories of these objects. 

The micrometeorite record may indicate a need for an additional source of hydrated material in the outer solar system. Only about 10\% of micrometeorites come from asteroidal sources \citep{Nesvorny2010}, with the remaining material derived from cometary sources that are presumed to be anhydrous. Surveys suggest that about 20-30\% of micrometeorites are hydrated \citep[e.g.,][]{Genge2020,Keller2022}, much more than can be apparently produced from the fraction of hydrated asteroids in the main belt. Assessing the hydration state of various outer solar system populations can therefore establish whether there is sufficient material to explain this discrepancy between dynamical models and observations. While we provide evidence that irregular satellites' parent populations could provide such a source, further investigations to measure the overall fraction of hydrated irregular satellites are required to truly test this possibility.

We suggest that the exterior properties of irregular satellites may shed light on the interiors of Kuiper belt objects. For example, planetesimals in the outer solar system could have experienced enough heating for local aqueous alteration to occur, but their outer layers remained unaltered. In this case, fragmentation of such a body would produce hydrated and non-hydrated objects that both share the same origins. While speculative, this hypothesis is a relevant lens to consider when exploring the compositional variation amongst irregular satellites, or any highly collisionally evolved population.

\begin{acknowledgments}
\section{Acknowledgments}
The authors wish to thank David Nesvorn\'{y} and David Vokrouhlick\'{y} for useful conversations regarding the evolution of objects discussed in this paper, and John Noonan and Theodore Kareta for discussions regarding data processing and presentation.

We also thank two anonymous reviewers for their helpful feedback, which substantively improved this manuscript.

This work was supported by a NASA Earth and Space Science Fellowship (PI: Sharkey). We thank the IRTF telescope operators and MKSS staff for their support. The authors wish to recognize and acknowledge the significant cultural role and reverence the summit of Mauna Kea has always had within the indigenous Hawaiian community. We are most fortunate to have the opportunity to conduct observations from this mountain.

This work uses data taken with the MODS spectrographs built with funding from NSF grant AST-9987045 and the NSF Telescope System Instrumentation Program (TSIP), with additional funds from the Ohio Board of Regents and the Ohio State University Office of Research. The LBT is an international collaboration among institutions in the United States, Italy, and Germany. LBT Corporation partners are: The University of Arizona on behalf of the Arizona Board of Regents; Istituto Nazionale di Astrofisica, Italy; LBT Beteiligungsgesellschaft, Germany, representing the Max-Planck Society, The Leibniz Institute for Astrophysics Potsdam, and Heidelberg University; The Ohio State University, representing OSU, University of Notre Dame, University of Minnesota and University of Virginia.
\end{acknowledgments}
 \bibliography{bibliography}

\begin{thebibliography}{}
\expandafter\ifx\csname natexlab\endcsname\relax\def\natexlab#1{#1}\fi
\providecommand{\url}[1]{\href{#1}{#1}}
\providecommand{\dodoi}[1]{doi:~\href{http://doi.org/#1}{\nolinkurl{#1}}}
\providecommand{\doeprint}[1]{\href{http://ascl.net/#1}{\nolinkurl{http://ascl.net/#1}}}
\providecommand{\doarXiv}[1]{\href{https://arxiv.org/abs/#1}{\nolinkurl{https://arxiv.org/abs/#1}}}

\bibitem[{{Astakhov} {et~al.}(2003){Astakhov}, {Burbanks}, {Wiggins}, \&
  {Farrelly}}]{Astakhov2003}
{Astakhov}, S.~A., {Burbanks}, A.~D., {Wiggins}, S., \& {Farrelly}, D. 2003,
  \nat, 423, 264, \dodoi{10.1038/nature01622}

\bibitem[{{Barucci} {et~al.}(2005){Barucci}, {Belskaya}, {Fulchignoni}, \&
  {Birlan}}]{Barucci2005}
{Barucci}, M.~A., {Belskaya}, I.~N., {Fulchignoni}, M., \& {Birlan}, M. 2005,
  \aj, 130, 1291, \dodoi{10.1086/431957}

\bibitem[{{Belli} {et~al.}(2018){Belli}, {Contursi}, \& {Davies}}]{Belli2018}
{Belli}, S., {Contursi}, A., \& {Davies}, R.~I. 2018, \mnras, 478, 2097,
  \dodoi{10.1093/mnras/sty1236}

\bibitem[{{Benecchi} {et~al.}(2010){Benecchi}, {Noll}, {Grundy}, \&
  {Levison}}]{Benecchi2010}
{Benecchi}, S.~D., {Noll}, K.~S., {Grundy}, W.~M., \& {Levison}, H.~F. 2010,
  \icarus, 207, 978, \dodoi{10.1016/j.icarus.2009.12.017}

\bibitem[{{Bhatt} {et~al.}(2017){Bhatt}, {Reddy}, {Schindler}, {Cloutis},
  {Bhardwaj}, {Corre}, \& {Mann}}]{Bhatt2017}
{Bhatt}, M., {Reddy}, V., {Schindler}, K., {et~al.} 2017, \aap, 608, A67,
  \dodoi{10.1051/0004-6361/201630361}

\bibitem[{{Bierson} \& {Nimmo}(2019)}]{Bierson2019}
{Bierson}, C.~J., \& {Nimmo}, F. 2019, \icarus, 326, 10,
  \dodoi{10.1016/j.icarus.2019.01.027}

\bibitem[{{Bohlin} {et~al.}(2014){Bohlin}, {Gordon}, \&
  {Tremblay}}]{Bohlin2014}
{Bohlin}, R.~C., {Gordon}, K.~D., \& {Tremblay}, P.~E. 2014, \pasp, 126, 711,
  \dodoi{10.1086/677655}

\bibitem[{{Bottke} {et~al.}(2010){Bottke}, {Nesvorn{\'y}}, {Vokrouhlick{\'y}},
  \& {Morbidelli}}]{Bottke2010}
{Bottke}, W.~F., {Nesvorn{\'y}}, D., {Vokrouhlick{\'y}}, D., \& {Morbidelli},
  A. 2010, \aj, 139, 994, \dodoi{10.1088/0004-6256/139/3/994}

\bibitem[{{Bottke} {et~al.}(2013){Bottke}, {Vokrouhlick{\'y}}, {Nesvorn{\'y}},
  \& {Moore}}]{Bottke2013}
{Bottke}, W.~F., {Vokrouhlick{\'y}}, D., {Nesvorn{\'y}}, D., \& {Moore}, J.~M.
  2013, \icarus, 223, 775, \dodoi{10.1016/j.icarus.2013.01.008}

\bibitem[{{Brown}(2000)}]{mebrown2000}
{Brown}, M.~E. 2000, \aj, 119, 977, \dodoi{10.1086/301202}

\bibitem[{{Brown} {et~al.}(1998){Brown}, {Koresko}, \& {Blake}}]{mebrown1998}
{Brown}, M.~E., {Koresko}, C.~D., \& {Blake}, G.~A. 1998, \apjl, 508, L175,
  \dodoi{10.1086/311741}

\bibitem[{{Brown} \& {Rhoden}(2014)}]{Brown2014}
{Brown}, M.~E., \& {Rhoden}, A.~R. 2014, \apjl, 793, L44,
  \dodoi{10.1088/2041-8205/793/2/L44}

\bibitem[{{Brown} {et~al.}(1999){Brown}, {Cruikshank}, {Pendleton}, \&
  {Veeder}}]{rhbrown1999}
{Brown}, R.~H., {Cruikshank}, D.~P., {Pendleton}, Y., \& {Veeder}, G.~J. 1999,
  \icarus, 139, 374, \dodoi{10.1006/icar.1999.6109}

\bibitem[{{Brozovi{\'c}} \& {Jacobson}(2017)}]{Brozovic2017}
{Brozovi{\'c}}, M., \& {Jacobson}, R.~A. 2017, \aj, 153, 147,
  \dodoi{10.3847/1538-3881/aa5e4d}

\bibitem[{{Buratti} {et~al.}(1997){Buratti}, {Goguen}, \&
  {Mosher}}]{buratti1997}
{Buratti}, B.~J., {Goguen}, J.~D., \& {Mosher}, J.~A. 1997, \icarus, 126, 225,
  \dodoi{10.1006/icar.1996.5644}

\bibitem[{{Carry} {et~al.}(2021){Carry}, {Vernazza}, {Vachier}, {Neveu},
  {Berthier}, {Hanu{\v{s}}}, {Ferrais}, {Jorda}, {Marsset}, {Viikinkoski},
  {Bartczak}, {Behrend}, {Benkhaldoun}, {Birlan}, {Castillo-Rogez}, {Cipriani},
  {Colas}, {Drouard}, {Dudzi{\'n}ski}, {Desmars}, {Dumas}, {{\v{D}}urech},
  {Fetick}, {Fusco}, {Grice}, {Jehin}, {Kaasalainen}, {Kryszczynska}, {Lamy},
  {Marchis}, {Marciniak}, {Michalowski}, {Michel}, {Pajuelo}, {Podlewska-Gaca},
  {Rambaux}, {Santana-Ros}, {Storrs}, {Tanga}, {Vigan}, {Warner}, {Wieczorek},
  {Witasse}, \& {Yang}}]{Carry2021}
{Carry}, B., {Vernazza}, P., {Vachier}, F., {et~al.} 2021, \aap, 650, A129,
  \dodoi{10.1051/0004-6361/202140342}

\bibitem[{{Cartwright} {et~al.}(2018){Cartwright}, {Emery}, {Pinilla-Alonso},
  {Lucas}, {Rivkin}, \& {Trilling}}]{cartwright2018}
{Cartwright}, R.~J., {Emery}, J.~P., {Pinilla-Alonso}, N., {et~al.} 2018,
  \icarus, 314, 210, \dodoi{10.1016/j.icarus.2018.06.004}

\bibitem[{{Cartwright} {et~al.}(2020){Cartwright}, {Nordheim}, {Cruikshank},
  {Hand}, {Roser}, {Grundy}, {Beddingfield}, \& {Emery}}]{Cartwright2020}
{Cartwright}, R.~J., {Nordheim}, T.~A., {Cruikshank}, D.~P., {et~al.} 2020,
  \apjl, 902, L38, \dodoi{10.3847/2041-8213/abbdae}

\bibitem[{{Castillo-Rogez} {et~al.}(2012){Castillo-Rogez}, {Johnson}, {Thomas},
  {Choukroun}, {Matson}, \& {Lunine}}]{Castillo-Rogez2012}
{Castillo-Rogez}, J.~C., {Johnson}, T.~V., {Thomas}, P.~C., {et~al.} 2012,
  \icarus, 219, 86, \dodoi{10.1016/j.icarus.2012.02.002}

\bibitem[{{Clark} {et~al.}(2005){Clark}, {Brown}, {Jaumann}, {Cruikshank},
  {Nelson}, {Buratti}, {McCord}, {Lunine}, {Baines}, {Bellucci}, {Bibring},
  {Capaccioni}, {Cerroni}, {Coradini}, {Formisano}, {Langevin}, {Matson},
  {Mennella}, {Nicholson}, {Sicardy}, {Sotin}, {Hoefen}, {Curchin}, {Hansen},
  {Hibbitts}, \& {Matz}}]{Clark2005}
{Clark}, R.~N., {Brown}, R.~H., {Jaumann}, R., {et~al.} 2005, \nat, 435, 66,
  \dodoi{10.1038/nature03558}

\bibitem[{{Cloutis} {et~al.}(2011){Cloutis}, {Hiroi}, {Gaffey}, {Alexander}, \&
  {Mann}}]{Cloutis2011a}
{Cloutis}, E.~A., {Hiroi}, T., {Gaffey}, M.~J., {Alexander}, C.~M.~O.~D., \&
  {Mann}, P. 2011, \icarus, 212, 180, \dodoi{10.1016/j.icarus.2010.12.009}

\bibitem[{{Cloutis} {et~al.}(2018){Cloutis}, {Pietrasz}, {Kiddell}, {Izawa},
  {Vernazza}, {Burbine}, {DeMeo}, {Tait}, {Bell}, {Mann}, {Applin}, \&
  {Reddy}}]{Cloutis2018}
{Cloutis}, E.~A., {Pietrasz}, V.~B., {Kiddell}, C., {et~al.} 2018, \icarus,
  305, 203, \dodoi{10.1016/j.icarus.2018.01.015}

\bibitem[{{Colombo} \& {Franklin}(1971)}]{Colombo1971}
{Colombo}, G., \& {Franklin}, F.~A. 1971, \icarus, 15, 186,
  \dodoi{10.1016/0019-1035(71)90073-X}

\bibitem[{{Cruikshank}(1977)}]{Cruikshank1977}
{Cruikshank}, D.~P. 1977, \icarus, 30, 224,
  \dodoi{10.1016/0019-1035(77)90136-1}

\bibitem[{{{\'C}uk} \& {Burns}(2004)}]{Cuk2004}
{{\'C}uk}, M., \& {Burns}, J.~A. 2004, \icarus, 167, 369,
  \dodoi{10.1016/j.icarus.2003.09.026}

\bibitem[{{Cushing} {et~al.}(2004){Cushing}, {Vacca}, \&
  {Rayner}}]{cushing2004}
{Cushing}, M.~C., {Vacca}, W.~D., \& {Rayner}, J.~T. 2004, \pasp, 116, 362,
  \dodoi{10.1086/382907}

\bibitem[{{DiSanti} {et~al.}(2017){DiSanti}, {Bonev}, {Russo}, {Vervack},
  {Gibb}, {Roth}, {McKay}, {Kawakita}, {Feaga}, \& {Weaver}}]{DiSanti2017}
{DiSanti}, M.~A., {Bonev}, B.~P., {Russo}, N.~D., {et~al.} 2017, \aj, 154, 246,
  \dodoi{10.3847/1538-3881/aa8639}

\bibitem[{{Emery} \& {Brown}(2004)}]{emery2004}
{Emery}, J.~P., \& {Brown}, R.~H. 2004, \icarus, 170, 131,
  \dodoi{10.1016/j.icarus.2004.02.004}

\bibitem[{{Emery} {et~al.}(2011){Emery}, {Burr}, \& {Cruikshank}}]{emery2011}
{Emery}, J.~P., {Burr}, D.~M., \& {Cruikshank}, D.~P. 2011, \aj, 141, 25,
  \dodoi{10.1088/0004-6256/141/1/25}

\bibitem[{{Emery} {et~al.}(2006){Emery}, {Cruikshank}, \& {Van
  Cleve}}]{Emery2006}
{Emery}, J.~P., {Cruikshank}, D.~P., \& {Van Cleve}, J. 2006, \icarus, 182,
  496, \dodoi{10.1016/j.icarus.2006.01.011}

\bibitem[{{Farkas-Tak{\'a}cs} {et~al.}(2017){Farkas-Tak{\'a}cs}, {Kiss},
  {P{\'a}l}, {Moln{\'a}r}, {Szab{\'o}}, {Hanyecz}, {S{\'a}rneczky},
  {Szab{\'o}}, {Marton}, {Mommert}, {Szak{\'a}ts}, {M{\"u}ller}, \&
  {Kiss}}]{FarkasTakacs2017}
{Farkas-Tak{\'a}cs}, A., {Kiss}, C., {P{\'a}l}, A., {et~al.} 2017, \aj, 154,
  119, \dodoi{10.3847/1538-3881/aa8365}

\bibitem[{{Fornasier} {et~al.}(2007){Fornasier}, {Dotto}, {Hainaut}, {Marzari},
  {Boehnhardt}, {De Luise}, \& {Barucci}}]{Fornasier2007}
{Fornasier}, S., {Dotto}, E., {Hainaut}, O., {et~al.} 2007, \icarus, 190, 622,
  \dodoi{10.1016/j.icarus.2007.03.033}

\bibitem[{{Fraser} \& {Brown}(2018)}]{Fraser2018}
{Fraser}, W.~C., \& {Brown}, M.~E. 2018, \aj, 156, 23,
  \dodoi{10.3847/1538-3881/aac213}

\bibitem[{{Gaffey}(2010)}]{Gaffey2010}
{Gaffey}, M.~J. 2010, \icarus, 209, 564, \dodoi{10.1016/j.icarus.2010.05.006}

\bibitem[{{Genge} {et~al.}(2020){Genge}, {Van Ginneken}, \&
  {Suttle}}]{Genge2020}
{Genge}, M.~J., {Van Ginneken}, M., \& {Suttle}, M.~D. 2020, \planss, 187,
  104900, \dodoi{10.1016/j.pss.2020.104900}

\bibitem[{{Grav} \& {Bauer}(2007)}]{Grav2007}
{Grav}, T., \& {Bauer}, J. 2007, \icarus, 191, 267,
  \dodoi{10.1016/j.icarus.2007.04.020}

\bibitem[{{Grav} {et~al.}(2015){Grav}, {Bauer}, {Mainzer}, {Masiero}, {Nugent},
  {Cutri}, {Sonnett}, \& {Kramer}}]{Grav2015}
{Grav}, T., {Bauer}, J.~M., {Mainzer}, A.~K., {et~al.} 2015, \apj, 809, 3,
  \dodoi{10.1088/0004-637X/809/1/3}

\bibitem[{{Grav} \& {Holman}(2004)}]{Grav2004a}
{Grav}, T., \& {Holman}, M.~J. 2004, \apjl, 605, L141, \dodoi{10.1086/420881}

\bibitem[{{Grav} {et~al.}(2004){Grav}, {Holman}, \& {Fraser}}]{Grav2004b}
{Grav}, T., {Holman}, M.~J., \& {Fraser}, W.~C. 2004, \apjl, 613, L77,
  \dodoi{10.1086/424997}

\bibitem[{{Grav} {et~al.}(2003){Grav}, {Holman}, {Gladman}, \&
  {Aksnes}}]{Grav2003}
{Grav}, T., {Holman}, M.~J., {Gladman}, B.~J., \& {Aksnes}, K. 2003, \icarus,
  166, 33, \dodoi{10.1016/j.icarus.2003.07.005}

\bibitem[{{Grav} {et~al.}(2012){Grav}, {Mainzer}, {Bauer}, {Masiero}, \&
  {Nugent}}]{Grav2012}
{Grav}, T., {Mainzer}, A.~K., {Bauer}, J.~M., {Masiero}, J.~R., \& {Nugent},
  C.~R. 2012, \apj, 759, 49, \dodoi{10.1088/0004-637X/759/1/49}

\bibitem[{{Graykowski} \& {Jewitt}(2018)}]{Graykowski2018}
{Graykowski}, A., \& {Jewitt}, D. 2018, \aj, 155, 184,
  \dodoi{10.3847/1538-3881/aab49b}

\bibitem[{{Grundy} {et~al.}(2007){Grundy}, {Stansberry}, {Noll}, {Stephens},
  {Trilling}, {Kern}, {Spencer}, {Cruikshank}, \& {Levison}}]{Grundy2007}
{Grundy}, W.~M., {Stansberry}, J.~A., {Noll}, K.~S., {et~al.} 2007, \icarus,
  191, 286, \dodoi{10.1016/j.icarus.2007.04.004}

\bibitem[{{Heppenheimer} \& {Porco}(1977)}]{Heppenheimer1977}
{Heppenheimer}, T.~A., \& {Porco}, C. 1977, \icarus, 30, 385,
  \dodoi{10.1016/0019-1035(77)90173-7}

\bibitem[{Hopp {et~al.}(2022)Hopp, Dauphas, Abe, Aléon, Alexander, Amari,
  Amelin, ichi Bajo, Bizzarro, Bouvier, Carlson, Chaussidon, Choi, Davis,
  Rocco, Fujiya, Fukai, Gautam, Haba, Hibiya, Hidaka, Homma, Hoppe, Huss,
  Ichida, Iizuka, Ireland, Ishikawa, Ito, Itoh, Kawasaki, Kita, Kitajima,
  Kleine, Komatani, Krot, Liu, Masuda, McKeegan, Morita, Motomura, Moynier,
  Nakai, Nagashima, Nesvorný, Nguyen, Nittler, Onose, Pack, Park, Piani, Qin,
  Russell, Sakamoto, Schönbächler, Tafla, Tang, Terada, Terada, Usui, Wada,
  Wadhwa, Walker, Yamashita, Yin, Yokoyama, Yoneda, Young, Yui, Zhang,
  Nakamura, Naraoka, Noguchi, Okazaki, Sakamoto, Yabuta, Abe, Miyazaki, Nakato,
  Nishimura, Okada, Yada, Yogata, Nakazawa, Saiki, Tanaka, Terui, Tsuda, ichiro
  Watanabe, Yoshikawa, Tachibana, \& Yurimoto}]{Hopp2022}
Hopp, T., Dauphas, N., Abe, Y., {et~al.} 2022, Science Advances, 0, eadd8141,
  \dodoi{10.1126/sciadv.add8141}

\bibitem[{{Horne}(1986)}]{Horne1986}
{Horne}, K. 1986, \pasp, 98, 609, \dodoi{10.1086/131801}

\bibitem[{{Howell} {et~al.}(2011){Howell}, {Rivkin}, {Vilas}, {Magri}, {Nolan},
  {Vervack}, \& {Fernandez}}]{Howell2011}
{Howell}, E.~S., {Rivkin}, A.~S., {Vilas}, F., {et~al.} 2011, in EPSC-DPS Joint
  Meeting 2011, Vol. 2011, 637

\bibitem[{{Humes} {et~al.}(2022){Humes}, {Thomas}, {Emery}, \&
  {Grundy}}]{Humes2022}
{Humes}, O.~A., {Thomas}, C.~A., {Emery}, J.~P., \& {Grundy}, W.~M. 2022, \psj,
  3, 190, \dodoi{10.3847/PSJ/ac8059}

\bibitem[{{Jarvis} {et~al.}(2000){Jarvis}, {Vilas}, {Larson}, \&
  {Gaffey}}]{Jarvis2000}
{Jarvis}, K.~S., {Vilas}, F., {Larson}, S.~M., \& {Gaffey}, M.~J. 2000,
  \icarus, 145, 445, \dodoi{10.1006/icar.2000.6344}

\bibitem[{{Jewitt}(2018)}]{Jewitt2018}
{Jewitt}, D. 2018, \aj, 155, 56, \dodoi{10.3847/1538-3881/aaa1a4}

\bibitem[{{Jewitt} \& {Haghighipour}(2007)}]{Jewitt2007}
{Jewitt}, D., \& {Haghighipour}, N. 2007, \araa, 45, 261,
  \dodoi{10.1146/annurev.astro.44.051905.092459}

\bibitem[{{Karkoschka}(1997)}]{karkoschka1997}
{Karkoschka}, E. 1997, \icarus, 125, 348, \dodoi{10.1006/icar.1996.5631}

\bibitem[{{Keller} \& {Flynn}(2022)}]{Keller2022}
{Keller}, L.~P., \& {Flynn}, G.~J. 2022, Nature Astronomy, 6, 731,
  \dodoi{10.1038/s41550-022-01647-6}

\bibitem[{{Kelson}(2003)}]{Kelson2003}
{Kelson}, D.~D. 2003, \pasp, 115, 688, \dodoi{10.1086/375502}

\bibitem[{{King} \& {Clark}(1989)}]{King1989}
{King}, T. V.~V., \& {Clark}, R.~N. 1989, \jgr, 94, 13997,
  \dodoi{10.1029/JB094iB10p13997}

\bibitem[{{King} \& {Clark}(1997)}]{King1997}
{King}, T. V.~V., \& {Clark}, R.~N. 1997, in Lunar and Planetary Science
  Conference, Lunar and Planetary Science Conference, 727

\bibitem[{{Kiss} {et~al.}(2016){Kiss}, {P{\'a}l}, {Farkas-Tak{\'a}cs},
  {Szab{\'o}}, {Szab{\'o}}, {Kiss}, {Moln{\'a}r}, {S{\'a}rneczky},
  {M{\"u}ller}, {Mommert}, \& {Stansberry}}]{Kiss2016}
{Kiss}, C., {P{\'a}l}, A., {Farkas-Tak{\'a}cs}, A.~I., {et~al.} 2016, \mnras,
  457, 2908, \dodoi{10.1093/mnras/stw081}

\bibitem[{{Laczniak} {et~al.}(2021){Laczniak}, {Thompson}, {Christoffersen},
  {Dukes}, {Clemett}, {Morris}, \& {Keller}}]{Laczniak2021}
{Laczniak}, D.~L., {Thompson}, M.~S., {Christoffersen}, R., {et~al.} 2021,
  \icarus, 364, 114479, \dodoi{10.1016/j.icarus.2021.114479}

\bibitem[{{Lantz} {et~al.}(2018){Lantz}, {Binzel}, \& {DeMeo}}]{Lantz2018}
{Lantz}, C., {Binzel}, R.~P., \& {DeMeo}, F.~E. 2018, \icarus, 302, 10,
  \dodoi{10.1016/j.icarus.2017.11.010}

\bibitem[{{Levison} {et~al.}(2021){Levison}, {Olkin}, {Noll}, {Marchi}, {Bell},
  {Bierhaus}, {Binzel}, {Bottke}, {Britt}, {Brown}, {Buie}, {Christensen},
  {Emery}, {Grundy}, {Hamilton}, {Howett}, {Mottola}, {P{\"a}tzold}, {Reuter},
  {Spencer}, {Statler}, {Stern}, {Sunshine}, {Weaver}, \& {Wong}}]{Levison2021}
{Levison}, H.~F., {Olkin}, C.~B., {Noll}, K.~S., {et~al.} 2021, \psj, 2, 171,
  \dodoi{10.3847/PSJ/abf840}

\bibitem[{{Mahjoub} {et~al.}(2016){Mahjoub}, {Poston}, {Hand}, {Brown},
  {Hodyss}, {Blacksberg}, {Eiler}, {Carlson}, {Ehlmann}, \&
  {Choukroun}}]{Mahjoub2016}
{Mahjoub}, A., {Poston}, M.~J., {Hand}, K.~P., {et~al.} 2016, \apj, 820, 141,
  \dodoi{10.3847/0004-637X/820/2/141}

\bibitem[{{Mahjoub} {et~al.}(2017){Mahjoub}, {Poston}, {Blacksberg}, {Eiler},
  {Brown}, {Ehlmann}, {Hodyss}, {Hand}, {Carlson}, \&
  {Choukroun}}]{Mahjoub2017}
{Mahjoub}, A., {Poston}, M.~J., {Blacksberg}, J., {et~al.} 2017, \apj, 846,
  148, \dodoi{10.3847/1538-4357/aa85e0}

\bibitem[{{Maris} {et~al.}(2001){Maris}, {Carraro}, {Cremonese}, \&
  {Fulle}}]{Maris2001}
{Maris}, M., {Carraro}, G., {Cremonese}, G., \& {Fulle}, M. 2001, \aj, 121,
  2800, \dodoi{10.1086/320378}

\bibitem[{Markwardt(2023)}]{Markwardt2023}
Markwardt, L. 2023, PhD thesis, \dodoi{10.7302/7253}

\bibitem[{{Markwardt} {et~al.}(2023){Markwardt}, {Wen Lin}, {Gerdes}, \&
  {Adams}}]{Markwardt2023b}
{Markwardt}, L., {Wen Lin}, H., {Gerdes}, D., \& {Adams}, F.~C. 2023, \psj, 4,
  135, \dodoi{10.3847/PSJ/ace528}

\bibitem[{{Marschall} {et~al.}(2022){Marschall}, {Nesvorn{\'y}}, {Deienno},
  {Wong}, {Levison}, \& {Bottke}}]{Marschall2022}
{Marschall}, R., {Nesvorn{\'y}}, D., {Deienno}, R., {et~al.} 2022, \aj, 164,
  167, \dodoi{10.3847/1538-3881/ac8d6b}

\bibitem[{{Martin} \& {Emery}(2023)}]{Martin2023}
{Martin}, A.~C., \& {Emery}, J.~P. 2023, \psj, 4, 153,
  \dodoi{10.3847/PSJ/aced0c}

\bibitem[{{Martin} {et~al.}(2022){Martin}, {Emery}, \& {Loeffler}}]{Martin2022}
{Martin}, A.~C., {Emery}, J.~P., \& {Loeffler}, M.~J. 2022, \icarus, 378,
  114921, \dodoi{10.1016/j.icarus.2022.114921}

\bibitem[{{Mitchell} {et~al.}(2020){Mitchell}, {Reddy}, {Sharkey}, {Sanchez},
  {Burbine}, {Le Corre}, \& {Thomas}}]{Mitchell2020}
{Mitchell}, A.~M., {Reddy}, V., {Sharkey}, B. N.~L., {et~al.} 2020, \icarus,
  336, 113426, \dodoi{10.1016/j.icarus.2019.113426}

\bibitem[{{Nesvorn{\'y}}(2018)}]{Nesvorny2018}
{Nesvorn{\'y}}, D. 2018, \araa, 56, 137,
  \dodoi{10.1146/annurev-astro-081817-052028}

\bibitem[{{Nesvorn{\'y}} {et~al.}(2004){Nesvorn{\'y}}, {Beaug{\'e}}, \&
  {Dones}}]{Nesvorny2004}
{Nesvorn{\'y}}, D., {Beaug{\'e}}, C., \& {Dones}, L. 2004, \aj, 127, 1768,
  \dodoi{10.1086/382099}

\bibitem[{{Nesvorn{\'y}} {et~al.}(2010){Nesvorn{\'y}}, {Jenniskens}, {Levison},
  {Bottke}, {Vokrouhlick{\'y}}, \& {Gounelle}}]{Nesvorny2010}
{Nesvorn{\'y}}, D., {Jenniskens}, P., {Levison}, H.~F., {et~al.} 2010, \apj,
  713, 816, \dodoi{10.1088/0004-637X/713/2/816}

\bibitem[{{Nesvorn{\'y}} {et~al.}(2021){Nesvorn{\'y}}, {Li}, {Simon}, {Youdin},
  {Richardson}, {Marschall}, \& {Grundy}}]{Nesvorny2021}
{Nesvorn{\'y}}, D., {Li}, R., {Simon}, J.~B., {et~al.} 2021, \psj, 2, 27,
  \dodoi{10.3847/PSJ/abd858}

\bibitem[{{Nesvorn{\'y}} \& {Morbidelli}(2012)}]{Nesvorny2012}
{Nesvorn{\'y}}, D., \& {Morbidelli}, A. 2012, \aj, 144, 117,
  \dodoi{10.1088/0004-6256/144/4/117}

\bibitem[{{Nesvorn{\'y}} {et~al.}(2018){Nesvorn{\'y}}, {Vokrouhlick{\'y}},
  {Bottke}, \& {Levison}}]{Nesvorny2018b}
{Nesvorn{\'y}}, D., {Vokrouhlick{\'y}}, D., {Bottke}, W.~F., \& {Levison},
  H.~F. 2018, Nature Astronomy, 2, 878, \dodoi{10.1038/s41550-018-0564-3}

\bibitem[{{Nesvorn{\'y}} {et~al.}(2020){Nesvorn{\'y}}, {Vokrouhlick{\'y}},
  {Bottke}, {Levison}, \& {Grundy}}]{Nesvorny2020}
{Nesvorn{\'y}}, D., {Vokrouhlick{\'y}}, D., {Bottke}, W.~F., {Levison}, H.~F.,
  \& {Grundy}, W.~M. 2020, \apjl, 893, L16, \dodoi{10.3847/2041-8213/ab8311}

\bibitem[{{Nesvorn{\'y}} {et~al.}(2014){Nesvorn{\'y}}, {Vokrouhlick{\'y}}, \&
  {Deienno}}]{Nesvorny2014}
{Nesvorn{\'y}}, D., {Vokrouhlick{\'y}}, D., \& {Deienno}, R. 2014, \apj, 784,
  22, \dodoi{10.1088/0004-637X/784/1/22}

\bibitem[{{Nesvorn{\'y}} {et~al.}(2007){Nesvorn{\'y}}, {Vokrouhlick{\'y}}, \&
  {Morbidelli}}]{Nesvorny2007}
{Nesvorn{\'y}}, D., {Vokrouhlick{\'y}}, D., \& {Morbidelli}, A. 2007, \aj, 133,
  1962, \dodoi{10.1086/512850}

\bibitem[{{Nesvorn{\'y}} {et~al.}(2013){Nesvorn{\'y}}, {Vokrouhlick{\'y}}, \&
  {Morbidelli}}]{Nesvorny2013}
---. 2013, \apj, 768, 45, \dodoi{10.1088/0004-637X/768/1/45}

\bibitem[{{Nesvorn{\'y}} {et~al.}(2019){Nesvorn{\'y}}, {Vokrouhlick{\'y}},
  {Stern}, {Davidsson}, {Bannister}, {Volk}, {Chen}, {Gladman}, {Kavelaars},
  {Petit}, {Gwyn}, \& {Alexandersen}}]{Nesvorny2019}
{Nesvorn{\'y}}, D., {Vokrouhlick{\'y}}, D., {Stern}, A.~S., {et~al.} 2019, \aj,
  158, 132, \dodoi{10.3847/1538-3881/ab3651}

\bibitem[{{Nicholson} {et~al.}(2008){Nicholson}, {Cuk}, {Sheppard}, {Nesvorny},
  \& {Johnson}}]{Nicholson2008}
{Nicholson}, P.~D., {Cuk}, M., {Sheppard}, S.~S., {Nesvorny}, D., \& {Johnson},
  T.~V. 2008, in The Solar System Beyond Neptune, ed. M.~A. {Barucci},
  H.~{Boehnhardt}, D.~P. {Cruikshank}, A.~{Morbidelli}, \& R.~{Dotson}, 411

\bibitem[{{Noll} {et~al.}(2020){Noll}, {Grundy}, {Nesvorn{\'y}}, \&
  {Thirouin}}]{Noll2020}
{Noll}, K., {Grundy}, W.~M., {Nesvorn{\'y}}, D., \& {Thirouin}, A. 2020, in The
  Trans-Neptunian Solar System, ed. D.~{Prialnik}, M.~A. {Barucci}, \&
  L.~{Young}, 201--224, \dodoi{10.1016/B978-0-12-816490-7.00009-6}

\bibitem[{{Paradis} {et~al.}(2019){Paradis}, {Moeckel}, {Tollefson}, \& {de
  Pater}}]{Paradis2019}
{Paradis}, S., {Moeckel}, C., {Tollefson}, J., \& {de Pater}, I. 2019, \aj,
  158, 178, \dodoi{10.3847/1538-3881/ab4264}

\bibitem[{{Parker} {et~al.}(2013){Parker}, {Buie}, {Osip}, {Gwyn}, {Holman},
  {Borncamp}, {Spencer}, {Benecchi}, {Binzel}, {DeMeo}, {Fabbro}, {Fuentes},
  {Gay}, {Kavelaars}, {McLeod}, {Petit}, {Sheppard}, {Stern}, {Tholen},
  {Trilling}, {Ragozzine}, {Wasserman}, \& {Ice Hunters}}]{Parker2013}
{Parker}, A.~H., {Buie}, M.~W., {Osip}, D.~J., {et~al.} 2013, \aj, 145, 96,
  \dodoi{10.1088/0004-6256/145/4/96}

\bibitem[{{Peixinho} {et~al.}(2020){Peixinho}, {Thirouin}, {Tegler}, {Di
  Sisto}, {Delsanti}, {Guilbert-Lepoutre}, \& {Bauer}}]{Peixinho2020}
{Peixinho}, N., {Thirouin}, A., {Tegler}, S.~C., {et~al.} 2020, in The
  Trans-Neptunian Solar System, ed. D.~{Prialnik}, M.~A. {Barucci}, \&
  L.~{Young}, 307--329, \dodoi{10.1016/B978-0-12-816490-7.00014-X}

\bibitem[{{Perna} {et~al.}(2010){Perna}, {Barucci}, {Fornasier}, {DeMeo},
  {Alvarez-Candal}, {Merlin}, {Dotto}, {Doressoundiram}, \& {de
  Bergh}}]{Perna2010}
{Perna}, D., {Barucci}, M.~A., {Fornasier}, S., {et~al.} 2010, \aap, 510, A53,
  \dodoi{10.1051/0004-6361/200913654}

\bibitem[{{Pogge} {et~al.}(2010){Pogge}, {Atwood}, {Brewer}, {Byard},
  {Derwent}, {Gonzalez}, {Martini}, {Mason}, {O'Brien}, {Osmer}, {Pappalardo},
  {Steinbrecher}, {Teiga}, \& {Zhelem}}]{Pogge2010}
{Pogge}, R.~W., {Atwood}, B., {Brewer}, D.~F., {et~al.} 2010, Society of
  Photo-Optical Instrumentation Engineers (SPIE) Conference Series, Vol. 7735,
  {The multi-object double spectrographs for the Large Binocular Telescope},
  77350A, \dodoi{10.1117/12.857215}

\bibitem[{{Pollack} {et~al.}(1979){Pollack}, {Burns}, \&
  {Tauber}}]{Pollack1979}
{Pollack}, J.~B., {Burns}, J.~A., \& {Tauber}, M.~E. 1979, \icarus, 37, 587,
  \dodoi{10.1016/0019-1035(79)90016-2}

\bibitem[{{Poston} {et~al.}(2018){Poston}, {Mahjoub}, {Ehlmann}, {Blacksberg},
  {Brown}, {Carlson}, {Eiler}, {Hand}, {Hodyss}, \& {Wong}}]{Poston2018}
{Poston}, M.~J., {Mahjoub}, A., {Ehlmann}, B.~L., {et~al.} 2018, \apj, 856,
  124, \dodoi{10.3847/1538-4357/aab1f1}

\bibitem[{{Rettig} {et~al.}(2001){Rettig}, {Walsh}, \&
  {Consolmagno}}]{Rettig2001}
{Rettig}, T.~W., {Walsh}, K., \& {Consolmagno}, G. 2001, \icarus, 154, 313,
  \dodoi{10.1006/icar.2001.6715}

\bibitem[{{Roig} {et~al.}(2008){Roig}, {Ribeiro}, \& {Gil-Hutton}}]{Roig2008}
{Roig}, F., {Ribeiro}, A.~O., \& {Gil-Hutton}, R. 2008, \aap, 483, 911,
  \dodoi{10.1051/0004-6361:20079177}

\bibitem[{{Romon} {et~al.}(2001){Romon}, {de Bergh}, {Barucci},
  {Doressoundiram}, {Cuby}, {Le Bras}, {Dout{\'e}}, \& {Schmitt}}]{Romon2001}
{Romon}, J., {de Bergh}, C., {Barucci}, M.~A., {et~al.} 2001, \aap, 376, 310,
  \dodoi{10.1051/0004-6361:20010934}

\bibitem[{{Sanchez} {et~al.}(2013){Sanchez}, {Michelsen}, {Reddy}, \&
  {Nathues}}]{sanchez2013}
{Sanchez}, J.~A., {Michelsen}, R., {Reddy}, V., \& {Nathues}, A. 2013, \icarus,
  225, 131, \dodoi{10.1016/j.icarus.2013.02.036}

\bibitem[{{Sanchez} {et~al.}(2015){Sanchez}, {Reddy}, {Dykhuis}, {Lindsay}, \&
  {Le Corre}}]{sanchez2015}
{Sanchez}, J.~A., {Reddy}, V., {Dykhuis}, M., {Lindsay}, S., \& {Le Corre}, L.
  2015, \apj, 808, 93, \dodoi{10.1088/0004-637X/808/1/93}

\bibitem[{{Seccull} {et~al.}(2018){Seccull}, {Fraser}, {Puzia}, {Brown}, \&
  {Sch{\"o}nebeck}}]{Seccull2018}
{Seccull}, T., {Fraser}, W.~C., {Puzia}, T.~H., {Brown}, M.~E., \&
  {Sch{\"o}nebeck}, F. 2018, \apjl, 855, L26, \dodoi{10.3847/2041-8213/aab3dc}

\bibitem[{{Sharkey} {et~al.}(2019){Sharkey}, {Reddy}, {Sanchez}, {Izawa}, \&
  {Emery}}]{Sharkey2019}
{Sharkey}, B. N.~L., {Reddy}, V., {Sanchez}, J.~A., {Izawa}, M. R.~M., \&
  {Emery}, J.~P. 2019, \aj, 158, 204, \dodoi{10.3847/1538-3881/ab46c0}

\bibitem[{{Sharkey} {et~al.}(2021{\natexlab{a}}){Sharkey}, {Reddy}, {Sanchez},
  {Izawa}, \& {Harris}}]{Sharkey2021a}
{Sharkey}, B. N.~L., {Reddy}, V., {Sanchez}, J.~A., {Izawa}, M. R.~M., \&
  {Harris}, W.~M. 2021{\natexlab{a}}, \psj, 2, 143, \dodoi{10.3847/PSJ/ac0bbe}

\bibitem[{{Sharkey} {et~al.}(2021{\natexlab{b}}){Sharkey}, {Reddy}, {Malhotra},
  {Thirouin}, {Kuhn}, {Conrad}, {Rothberg}, {Sanchez}, {Thompson}, \&
  {Veillet}}]{Sharkey2021b}
{Sharkey}, B. N.~L., {Reddy}, V., {Malhotra}, R., {et~al.} 2021{\natexlab{b}},
  Communications Earth and Environment, 2, 231,
  \dodoi{10.1038/s43247-021-00303-7}

\bibitem[{{Souza-Feliciano} {et~al.}(2020){Souza-Feliciano}, {De Pr{\'a}},
  {Pinilla-Alonso}, {Alvarez-Candal}, {Fern{\'a}ndez-Valenzuela}, {De
  Le{\'o}n}, {Binzel}, {Arcoverde}, {Rond{\'o}n}, \&
  {Evangelista}}]{SouzaFeliciano2020}
{Souza-Feliciano}, A.~C., {De Pr{\'a}}, M., {Pinilla-Alonso}, N., {et~al.}
  2020, \icarus, 338, 113463, \dodoi{10.1016/j.icarus.2019.113463}

\bibitem[{{Stansberry} {et~al.}(2012){Stansberry}, {Grundy}, {Mueller},
  {Benecchi}, {Rieke}, {Noll}, {Buie}, {Levison}, {Porter}, \&
  {Roe}}]{Stansberry2012}
{Stansberry}, J.~A., {Grundy}, W.~M., {Mueller}, M., {et~al.} 2012, \icarus,
  219, 676, \dodoi{10.1016/j.icarus.2012.03.029}

\bibitem[{{Sykes} {et~al.}(2000){Sykes}, {Nelson}, {Cutri}, {Kirkpatrick},
  {Hurt}, \& {Skrutskie}}]{Sykes2000}
{Sykes}, M.~V., {Nelson}, B., {Cutri}, R.~M., {et~al.} 2000, \icarus, 143, 371,
  \dodoi{10.1006/icar.1999.6269}

\bibitem[{{Szab{\'o}} {et~al.}(2007){Szab{\'o}}, {Ivezi{\'c}}, {Juri{\'c}}, \&
  {Lupton}}]{Szabo2007}
{Szab{\'o}}, G.~M., {Ivezi{\'c}}, {\v{Z}}., {Juri{\'c}}, M., \& {Lupton}, R.
  2007, \mnras, 377, 1393, \dodoi{10.1111/j.1365-2966.2007.11687.x}

\bibitem[{{Takir} \& {Emery}(2012)}]{Takir2012}
{Takir}, D., \& {Emery}, J.~P. 2012, \icarus, 219, 641,
  \dodoi{10.1016/j.icarus.2012.02.022}

\bibitem[{{Thomas} {et~al.}(2007){Thomas}, {Burns}, {Helfenstein}, {Squyres},
  {Veverka}, {Porco}, {Turtle}, {McEwen}, {Denk}, {Giese}, {Roatsch},
  {Johnson}, \& {Jacobson}}]{Thomas2007}
{Thomas}, P.~C., {Burns}, J.~A., {Helfenstein}, P., {et~al.} 2007, \icarus,
  190, 573, \dodoi{10.1016/j.icarus.2007.03.012}

\bibitem[{{Tsiganis} {et~al.}(2005){Tsiganis}, {Gomes}, {Morbidelli}, \&
  {Levison}}]{Tsiganis2005}
{Tsiganis}, K., {Gomes}, R., {Morbidelli}, A., \& {Levison}, H.~F. 2005, \nat,
  435, 459, \dodoi{10.1038/nature03539}

\bibitem[{{Vilas}(1994)}]{Vilas1994}
{Vilas}, F. 1994, \icarus, 111, 456, \dodoi{10.1006/icar.1994.1156}

\bibitem[{{Vilas} \& {Gaffey}(1989)}]{Vilas1989}
{Vilas}, F., \& {Gaffey}, M.~J. 1989, Science, 246, 790,
  \dodoi{10.1126/science.246.4931.790}

\bibitem[{{Vilas} \& {Hendrix}(2022)}]{Vilas2022}
{Vilas}, F., \& {Hendrix}, A. 2022, in AAS/Division for Planetary Sciences
  Meeting Abstracts, Vol.~54, AAS/Division for Planetary Sciences Meeting
  Abstracts, 308.04

\bibitem[{{Vilas} {et~al.}(2006){Vilas}, {Lederer}, {Gill}, {Jarvis}, \&
  {Thomas-Osip}}]{Vilas2006}
{Vilas}, F., {Lederer}, S.~M., {Gill}, S.~L., {Jarvis}, K.~S., \&
  {Thomas-Osip}, J.~E. 2006, \icarus, 180, 453,
  \dodoi{10.1016/j.icarus.2005.10.004}

\bibitem[{{Vilenius} {et~al.}(2012){Vilenius}, {Kiss}, {Mommert}, {M{\"u}ller},
  {Santos-Sanz}, {Pal}, {Stansberry}, {Mueller}, {Peixinho}, {Fornasier},
  {Lellouch}, {Delsanti}, {Thirouin}, {Ortiz}, {Duffard}, {Perna}, {Szalai},
  {Protopapa}, {Henry}, {Hestroffer}, {Rengel}, {Dotto}, \&
  {Hartogh}}]{Vilenius2012}
{Vilenius}, E., {Kiss}, C., {Mommert}, M., {et~al.} 2012, \aap, 541, A94,
  \dodoi{10.1051/0004-6361/201118743}

\bibitem[{{Vilenius} {et~al.}(2014){Vilenius}, {Kiss}, {M{\"u}ller}, {Mommert},
  {Santos-Sanz}, {P{\'a}l}, {Stansberry}, {Mueller}, {Peixinho}, {Lellouch},
  {Fornasier}, {Delsanti}, {Thirouin}, {Ortiz}, {Duffard}, {Perna}, \&
  {Henry}}]{Vilenius2014}
{Vilenius}, E., {Kiss}, C., {M{\"u}ller}, T., {et~al.} 2014, \aap, 564, A35,
  \dodoi{10.1051/0004-6361/201322416}

\bibitem[{{Vokrouhlick{\'y}} {et~al.}(2016){Vokrouhlick{\'y}}, {Bottke}, \&
  {Nesvorn{\'y}}}]{Vokrouhlicky2016}
{Vokrouhlick{\'y}}, D., {Bottke}, W.~F., \& {Nesvorn{\'y}}, D. 2016, \aj, 152,
  39, \dodoi{10.3847/0004-6256/152/2/39}

\bibitem[{{Wong} \& {Brown}(2016)}]{Wong2016}
{Wong}, I., \& {Brown}, M.~E. 2016, \aj, 152, 90,
  \dodoi{10.3847/0004-6256/152/4/90}

\bibitem[{{Wong} {et~al.}(2019){Wong}, {Brown}, {Blacksberg}, {Ehlmann}, \&
  {Mahjoub}}]{Wong2019}
{Wong}, I., {Brown}, M.~E., {Blacksberg}, J., {Ehlmann}, B.~L., \& {Mahjoub},
  A. 2019, \aj, 157, 161, \dodoi{10.3847/1538-3881/ab0e00}

\bibitem[{{Yang} \& {Jewitt}(2011)}]{Yang2011}
{Yang}, B., \& {Jewitt}, D. 2011, \aj, 141, 95,
  \dodoi{10.1088/0004-6256/141/3/95}

\bibitem[{{Yang} {et~al.}(2013){Yang}, {Lucey}, \& {Glotch}}]{Yang2013}
{Yang}, B., {Lucey}, P., \& {Glotch}, T. 2013, \icarus, 223, 359,
  \dodoi{10.1016/j.icarus.2012.11.025}

\end{thebibliography}

\begin{table}[p!]
\begin{center}
\resizebox{\textwidth}{!}{%
\begin{tabular}{|c|c|c|c|c|c|c|}
\hline \hline
\textbf{Object} & Pasiphae & Sinope & Lysithea & Ananke & Leda & Themisto \\
\hline
\textbf{Date} & 2021-06-19 & 2018-05-10 & 2018-05-21 & 2018-05-21 & 2018-04-19 & 2019-06-27  \\ 
              &            & 2018-05-13 &            & 2018-06-16 &            &             \\
\hline
\textbf{Instrument/Telescope} & SpeX/IRTF & SpeX/IRTF & SpeX/IRTF & SpeX/IRTF & MODS/LBT & MODS/LBT \\ 
\hline
\textbf{Visual}    & 17.7 & 18.2 & 18.3 & 18.3 & 20.0 & 20.1 \\
\textbf{Magnitude} &      & 18.3 &      & 18.6 &      &      \\
\hline
\textbf{Airmass}          & 1.2-1.6 & 1.4-1.6     & 1.3-1.4  & 1.2     & 1.5  & 1.7-1.8 \\
                          &         & 1.2-1.6 &      & 1.2-1.3 &      &      \\    
\hline 

\textbf{Seeing}          & 0.9"     & 1.2"    & 0.9"  & 0.9"     & 1.3"  & 1.0" \\
                          &         & 0.5" &      & 0.5" &      &      \\    
\hline 
\textbf{Phase}      & $10.2^{\circ}$ & $0.6^{\circ}$ & $2.3^{\circ}$ & $2.9^{\circ}$ & $3.9^{\circ}$  & $3.4^{\circ}$ \\
\textbf{Angle}      &                & $1.2^{\circ}$ &               & $7.1^{\circ}$ &                &      \\    
\hline                         
\textbf{Telluric}         & HD 211063 & SAO 158903 & SAO 158903 & SAO 158903 & Feige 34 & SAO 120107 \\
\textbf{Correction Stars} &           & SAO 158903 &            & SAO 158686 &          &            \\              
\hline
\textbf{Solar Analog Star}  & SAO 120107 & SAO 120107 & SAO 120107 & SAO 120107 & (*) & SAO 120107 \\
\hline
\textbf{Spectral Slope Baseline ($\mu m$)} & 1.00-2.00 & 1.00-2.00 & 1.00-2.00 & 1.00-2.00 & 0.61-0.80 & 0.62-0.80\\
\hline
\textbf{Spectral Slope ($\% / \mu m$ )} & $24 \pm 2$ & $31 \pm 1$ & $22 \pm 2$ & $31 \pm 4$ & $133 \pm 4$ & $-11 \pm 3$ \\
\hline

\end{tabular}}
\caption{Observational circumstances and derived spectral slopes for Jovian satellites characterized in this work. SpeX/IRTF observations were collected over the wavelength range of 0.75-2.55 microns, and MODS/LBT observations were collected over a wavelength range of 0.32-1.00 $\mu m$. HD 211063 is spectral type G2/3V D, and SAO 158903 is spectral type G0V C. Observations of Leda occurred with airmass variations $<0.05$. Quoted values of seeing for IRTF were taken on standard star focusing measurements prior to the beginning of target observations. For LBT observations, quoted seeing values were taken from the Guiding Control System measurements of the guide star. *Observations of Leda were interrupted before further calibration stars could be observed. Feige 34 was used to correct for telluric effects and to correct to solar colors. As noted in the text, spurious narrow-band artifacts may still be uncorrected in these spectra. Spectral slopes are computed from the final binned spectra as the flux ratios in a 0.1 $\mu m$-wide bin centered at either end of the spectral slope baseline values listed.}
\label{tab:jov_sats_tab}
\end{center}
\end{table}

\begin{table}[p!]
\begin{center}
\resizebox{\textwidth}{!}{%
\begin{tabular}{|c|c|c|c|c|c|c|}
\hline \hline
\textbf{Object} & Sycorax & Caliban & 2011 WG157 & 2010 TT191 & 2010 TS191 & 2006 RJ103 \\
\hline
\textbf{Type} & Uranian & Uranian & Neptune & Neptune & Neptune & Neptune \\
              & Irregular & Irregular & Trojan & Trojan & Trojan & Trojan \\
\hline
\textbf{Date} & 2018-11-15 & 2018-11-16 & 2018-11-16 & 2018-11-16 & 2018-11-15 & 2018-11-15  \\ 
\hline
\textbf{Instrument/} & MODS/LBT & MODS/LBT & MODS/LBT & MODS/LBT & MODS/LBT & MODS/LBT \\
\textbf{Telescope} &  &  &  &  &  &  \\
\hline
\textbf{Visual}    & 20.5 & 22.0 & 22.2 & 22.9 & 22.7 & 22.3 \\
\textbf{Magnitude} &      &      &      &      &      &      \\
\hline
\textbf{Airmass}   & 1.3-1.8 & 1.3-1.7 & 1.2-1.6  & 1.6-2.0 & 1.8-1.9 & 1.6-1.9 \\
\hline 
\textbf{Seeing} & 0.8" & 0.7" & 0.6"  & 0.8" & 0.9" & 0.8" \\
\hline                  
\textbf{Phase}      & $1.1^{\circ}$ & $1.2^{\circ}$ & $0.7^{\circ}$ & $0.5^{\circ}$ & $0.2^{\circ}$  & $0.4^{\circ}$ \\
\textbf{Angle}      &                &   &               &   &                &      \\    
\hline                         
\textbf{Telluric \&}       & HD 29714 & HD 29714 & HD 29714 & HD 29714 & HD 29714 & HD 29714 \\
\textbf{Solar Analog} &           &   &            &   &          &            \\              
\textbf{Star} &           &   &            &   &          &            \\  
\hline
\textbf{Spectral Slope Baseline ($\mu m$)} & 0.63-0.81 & 0.63-0.81 & 0.63-0.81 & 0.62-0.80 & 0.62-0.80 & 0.63-0.81\\
\hline
\textbf{Spectral Slope ($\% / \mu m$ )} & $130 \pm 6$ & $123 \pm 10$ & $75 \pm 8$ & $157 \pm 36$ & $110 \pm 22$ & $56 \pm 13$ \\
\hline

\end{tabular}}
\caption{Observational circumstances and derived spectral slopes for Uranian satellites and Neptune Trojans characterized in this work. MODS/LBT observations were collected over a wavelength range of 0.32-1.00 $\mu m$. For LBT observations, quoted seeing values were taken from the Guiding Control System measurements of the guide star. Spectral slopes are computed from the final binned spectra as the flux ratios in a 0.1 $\mu m$-wide bin centered at either end of the spectral slope baseline values listed.}
\label{tab:icegiant_obj_table}
\end{center}
\end{table}

\begin{table}[p!]
\begin{center}
\resizebox{\textwidth}{!}{%
\begin{tabular}{|c|c|c|c|c|}
\hline \hline
\textbf{Object} & Pasiphae & Sinope & Lysithea & Ananke \\
\hline
\textbf{Trojan Analog} & (3548) Eurybates & (11351) Leucus & (617) Patroclus & (617) Patroclus \\
\hline
\textbf{Spectral Classification} & C & D & P & P \\ 
\hline

\end{tabular}}
\caption{Summary of Jovian irregular satellites with infrared spectra collected in this work, and subsequent taxonomic classifications and analogous Jovian Trojan objects.}
\label{tab:irregs_trojans_comp}
\end{center}
\end{table}

\begin{table}[p!]
\begin{center}
\resizebox{\textwidth}{!}{%
\begin{tabular}{|c|c|c|c|c|c|c|}
\hline \hline
\textbf{Object} & \textbf{a (au)} & \textbf{e} & \textbf{I (deg)} & \textbf{Diameter} & \textbf{Bulk Density} & \textbf{Reference} \\
\hline
 (42355) Typhon–Echidna & 37.6 & 0.53 & 2.43 & $157 \pm 34$ km & $0.6^{+0.72}_{-0.29}\ g/cm^3$ & \citet{Stansberry2012} \\
\hline
 (65489) Ceto/Phorcys & 99.1 & 0.820	& 22.33 & $174 \pm 17$ km & $1.37^{+0.66}_{-0.30}\ g/cm^3$ & \citet{Grundy2007} \\ 
\hline
(88611) Teharonhiawako & 44.36 & 0.019 & 2.57 & $178^{+36}_{-33}$ km & $0.60^{+0.36}_{-0.33}\ g/cm^3$ &	\citet{Vilenius2014} \\
\hline
(612239) 2001 QC298 & 46.70	& 0.130	& 30.50	& $235^{+23}_{-21}$ km & $1.14^{+.034}_{-0.30}\ g/cm^3$ & \citet{Vilenius2014} \\
\hline
148780) Altjira & 44.42 & 0.062 & 5.19 & $123^{+23}_{-19}$ km & $0.3^{+0.50}_{-0.14}\ g/cm^3$ & \citet{Vilenius2014} \\
\hline
(79360) Sila-Nunam & 43.68  & 0.0065 & 2.25 & $249^{+31}_{-30}$ km & $0.73 \pm 0.28\ g/cm^3$ & \citet{Vilenius2012} \\
\hline
(47171) Lempo & 39.76 & 0.2316 & 8.412 & $286^{+45}_{-38}$ km & $0.542^{+0.317}_{-0.211}\ g/cm^3$ & \citet{Benecchi2010} \\
\hline
\end{tabular}}
\caption{Small ($D \lesssim 300$ km) TNOs with known bulk densities.}
\label{tab:TNO_densities}
\end{center}
\end{table}

\bibliographystyle{aasjournal}

\end{document}